\documentclass[aps,onecolumn,groupedaddress,superscriptaddress,amsmath,amssymb,amsthm]{revtex4}
\usepackage{subfigure,hyperref,bbm,times}
\usepackage{color}
\usepackage{graphicx}
\usepackage{dcolumn}
\usepackage{bm}
\newcommand{\bra}[1]{\langle#1|}
\newcommand{\ket}[1]{|#1\rangle}
\newcommand{\ave}[1]{\langle#1\rangle}

\providecommand{\openone}{\leavevmode\hbox{\small1\kern-3.8pt\normalsize1}}

\begin{document}

\title{Overview on the phenomenon of two-qubit entanglement revivals in classical environments}

\author{Rosario Lo Franco}
\email{rosario.lofranco@unipa.it}
\affiliation{Dipartimento di Energia, Ingegneria dell'Informazione e Modelli Matematici, Universit\`{a} di Palermo, Viale delle Scienze, Ed. 9, 90128 Palermo, Italy}
\affiliation{Dipartimento di Fisica e Chimica, Universit\`a di Palermo, via Archirafi 36, 90123 Palermo, Italy}
\author{Giuseppe Compagno}
\affiliation{Dipartimento di Fisica e Chimica, Universit\`a di Palermo, via Archirafi 36, 90123 Palermo, Italy}

\date{\today }

\begin{abstract}
The occurrence of revivals of quantum entanglement between separated open quantum systems has been shown not only for dissipative non-Markovian quantum environments but also for classical environments in absence of back-action. While the phenomenon is well understood in the first case, the possibility to retrieve entanglement when the composite quantum system is subject to local classical noise has generated a debate regarding its interpretation. This dynamical property of open quantum systems assumes an important role in quantum information theory from both fundamental and practical perspectives. Hybrid quantum-classical systems are in fact promising candidates to investigate the interplay among quantum and classical features and to look for possible control strategies of a quantum system by means of a classical device. 
Here we present an overview on this topic, reporting the most recent theoretical and experimental results about the revivals of entanglement between two qubits locally interacting with classical environments. We also review and discuss the interpretations provided so far to explain this phenomenon, suggesting that they can be cast under a unified viewpoint.  
\end{abstract}


\maketitle

\section{Introduction}
Quantum correlations, such as entanglement, nonlocality, steering and discord, among parts of composite systems are at the core of quantum theory and have also been acquiring a paramount importance as a resource for quantum information processes \cite{plenioreview,benenti,alickibook,horodecki2009RMP,petru,rivasreview,yu2009Science,lofrancoreview,aolitareview,Modi2012RMP,ABCreview,idpartSciRep,lofranco2006OpenSys,lofranco2009IJQI,lofranco2010PLA,amico2008RMP,mintertreview,laddNature,gisinQComm,vedral2014NatPhys}.  
Realistic systems are open and interact with the surrounding environment which usually has the effect to eventually destroy the quantum features of the system, even at a finite time, thus compromising their exploitation
\cite{yu2009Science,lofrancoreview,aolitareview,Modi2012RMP,xu2010NatComm,werlangPRA,sallesPRA,palermocatania2011IJQI,palermocatania2010PRA,kimble2007PRL,almeida2007Science}. Such a fate for quantum properties especially manifests within the configuration of independent qubits each one locally embedded in its own environment \cite{lofrancoreview}, which is the one required for implementing quantum communication and information protocols with distant individually addressable particles \cite{laddNature,gisinQComm}. Efforts are thus necessary to design efficient and feasible procedures to protect quantum correlations against detrimental noise. 

Under this perspective, in contrast to Markovian (memoryless) environments, suitable engineered environments capable to maintain quantum memory effects have been employed \cite{petru,rivasreview,yu2009Science,lofrancoreview,aolitareview,bellomo2007PRL,bellomo2008PRA,fonseca2012,lofrancoQIP,manPRA,manSciRep,bellomo2008trapping,bellomo2009ASL,haikkaPRA,gonzalez2016,lofrancoNJP,britoNJP,bylicka2014,AddisPRA,laurapseudo,non-Mar2,manNJP,PhysRevA.90.062104,Man:15,PhysRevA.93.042313,bellomo2010PhysScrManiscalco,bellomo2008bell,bellomoSavasta2011PhyScr,bellomo2011IJQI}. Such non-Markovian environments exhibit the general property to be necessary for revivals of quantum correlations to occur, irrespective of the fact whether they have either a quantum nature (e.g., a bosonic or fermionic environment) \cite{yu2009Science,lofrancoreview,aolitareview,bellomo2007PRL,bellomo2008PRA,xu2010PRL,bellomoSavasta2011PhyScr,fanchiniPRA} or a classical nature (e.g., stochastic noise, random field, phase noisy laser)\cite{zhou2010QIP,lofranco2012PhysScripta,bordone2012,lofranco2012PRA,LeggioPRA,LoFrancoNatCom,mannone2012,altintas2012PLA,benedetti2013,wilsonPRB,bellomo2012noisylaser,trapani2015,darrigo2013hidden,darrigo2014IJQI,darrigo2012AOP,adeline2014,PhysRevA.93.042119,benedettiIJQI,benedetti2013IEEE,PhysRevA.87.042310,rossiparis}. The possibility to have revivals of quantum correlations allow an extension of their exploitation time for some specific protocol. In order to make the revival phenomenon in open quantum systems easily reproducible and effective, it is of basic interest to understand its underlying mechanisms, particularly in light of the fact that it may happen under noise conditions originating from fundamentally different surrounding environments.

Revivals of entanglement between independent qubits after a finite time of complete disappearance have been first shown in the presence of non-Markovian dissipative quantum environments \cite{bellomo2007PRL,bellomo2008PRA}. Although the emergence of entanglement revivals under these conditions may appear strange at a first rapid look, it has been successively explained in terms of periodic entanglement exchanges among the qubits and the quantum constituents of the environment, because of the back-action of the local quantum environments on the qubits themselves allowed by the memory effects (see Fig.~1(a))\cite{bellomo2007PRL,Liu2011NatPhys,lopez2008PRL,Lopez2010PRA,chiuri2012,fanchiniSciRep}. On the other hand, the possibility to retrieve entanglement once it is destroyed between distant qubits locally subject to classical environments seems particularly counterintuitive, especially when such environments do not back react on the quantum system and are not able to store or share any quantum correlations. The first theoretical observations of entanglement revivals without environment back-action, for instance under random telegraph noise for solid state qubits \cite{zhou2010QIP,lofranco2012PhysScripta,bordone2012,lofrancoPRB,lofrancoecho}, put in evidence the importance of the phenomenon yet leaving open its interpretation. Closing this issue is not only relevant from a fundamental point of view regarding the classical-quantum border, but it also provides insights for the classical control of quantum systems with potential applications in future quantum technology requiring classical interfaces to operate  \cite{hybridPNAS2015,Milburn4469,Pavlovic,Altafini}. These considerations justify the wide interest in studying the evolution of quantum coherence and correlations in hybrid quantum-classical systems during the recent years \cite{zhou2010QIP,lofranco2012PhysScripta,bordone2012,lofranco2012PRA,LeggioPRA,LoFrancoNatCom,mannone2012,altintas2012PLA,benedetti2013,wilsonPRB,bellomo2012noisylaser,trapani2015,darrigo2013hidden,darrigo2014IJQI,darrigo2012AOP,adeline2014,PhysRevA.93.042119,benedettiIJQI,benedetti2013IEEE,PhysRevA.87.042310,rossiparis,Beggi2016,PhysRevA.93.042313,PhysRevA.89.032114,benedettiPLA,benedettiParisIJQI,Calvani23042013}. 

\begin{figure}[t]
\begin{center}
{\includegraphics[width=0.43\textwidth]{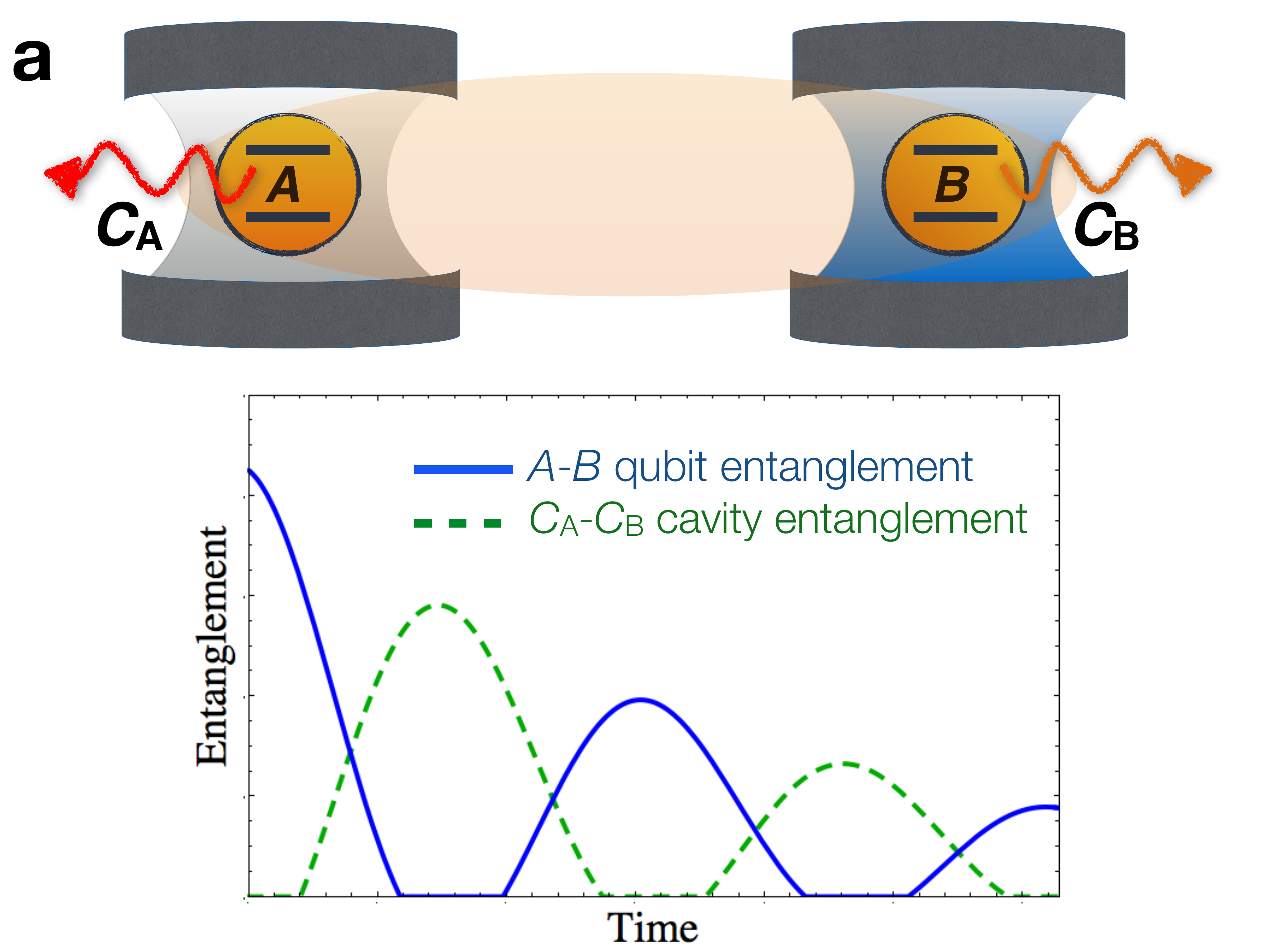}\hspace{1 cm}
\includegraphics[width=0.43\textwidth]{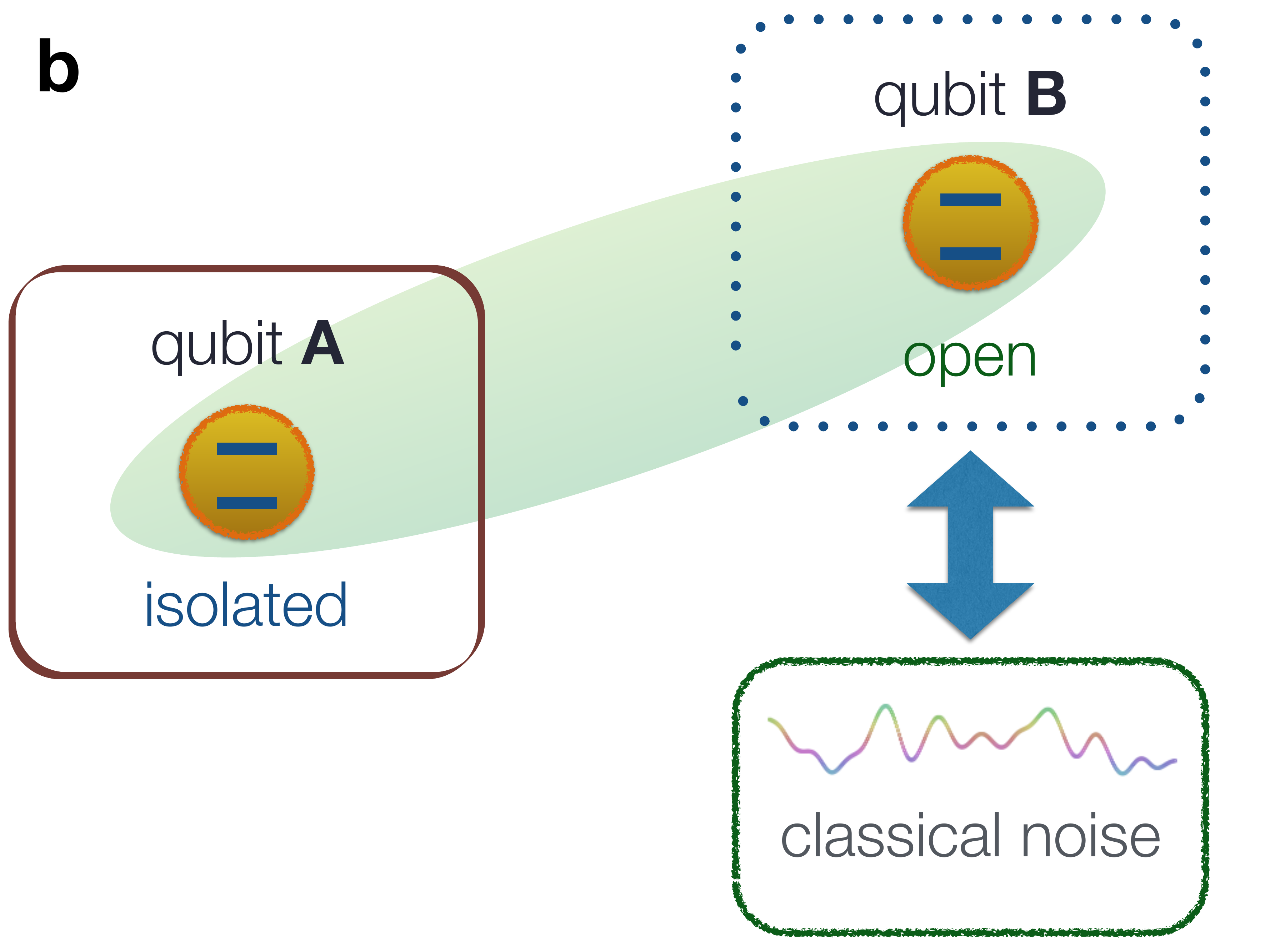}}
\end{center}
\caption{\textbf{Illustrations of the basic systems.} \textbf{a.} Two separated initially entangled qubits $A$ and $B$ locally interact with their own quantum environment represented by a cavity with high quality factor. The plot qualitatively show that the initial two-qubit entanglement spontaneously revive after being periodically transferred back and forth to the two cavities, thanks to the memory effects of the leaky cavities under non-Markovian conditions (see also Ref.~\cite{Lopez2010PRA}). \textbf{b.} A classical noise acts on the qubit $B$, whereas qubit $A$ is isolated. The two qubits are initially entangled. The two-qubit entanglement evolution under this configuration is qualitatively analogous to the one with both qubits locally interacting with their own environments.}
\label{fig:system}
\end{figure}

A convenient approach to understand the mechanisms underlying entanglement revivals in classical environments without back-action is to study simple feasible systems in order to minimize undesired side effects and make the role of classical noise prominent. The simplest possible open quantum system which fulfills this requirement is that depicted in Fig.~\ref{fig:system}(b), made of two initially entangled qubits, one of which (qubit $A$) is isolated, evolving according to its free Hamiltonian, while the second one (qubit $B$) interacts with a classical noise and thus evolves under the action of a non-unitary dynamical map. The two-qubit system and the classical environment are initially decoupled. Such a situation, which is paradigmatic for decoherence problems, is also known as the ``spectator configuration'' \cite{gonzalez2016}. This chapter is devoted to review the main theoretical results about the topic within this configuration, the experimental observations and the interpretations supplied so far. Moreover, the physical aspect that gathers the various interpretations under a unified framework is here provided.

In particular, the chapter is organized as follows. In Section~\ref{sec:TheorPredictions} we first discuss entanglement revivals in the case of  classical environment modeled by a random external field \cite{lofranco2012PRA,LeggioPRA}, which represents the first attempt to provide a simple model for deepening and understanding the phenomenon; we then also report two other models typical of the solid state where the classical noise is a low-frequency noise \cite{darrigo2012AOP} and a random telegraph noise (RTN) \cite{lofranco2012PhysScripta}. 
In Section~\ref{sec:ExpObserv} we describe two quantum optics experiments which reproduce the models with random external field \cite{LoFrancoNatCom} and low-frequency noise \cite{adeline2014}, respectively, and verify the existence of entanglement revivals. 
In Section~\ref{sec:Interp} we report the three known interpretations of the phenomenon of entanglement revivals in classical environments \cite{LoFrancoNatCom,LeggioPRA,darrigo2012AOP}, putting them under a unified view. 
We give our conclusions and outlook on the topic in Section~\ref{sec:conclusion}. 

\section{Theoretical predictions}\label{sec:TheorPredictions}

In this section we review the results about the revivals of entanglement between two qubits in the configuration of Fig.~\ref{fig:system}(b). We particularly focus on the case when the classical noise is simply modeled by a random external field and, successively, consider also the cases when the noise is the typical one encountered by superconducting qubits in the solid state such as longitudinal low-frequency noise and RTN. The two qubits are considered identical, that is with the same transition frequency $\omega_{0A}=\omega_{0B}=\omega_0$, and separated. The total Hamiltonian is thus in general given by $\mathcal{H}_\mathrm{tot}=\mathcal{H}_A+\mathcal{H}_B$, where $\mathcal{H}_A=-(\omega_0/2)\sigma_z$ is the free Hamiltonian of qubit $A$ where $\sigma_z = \ket{0}\bra{0} - \ket{1}\bra{1}$ is the third Pauli matrix. 
We shall see that the first two types of noise are capable to make entanglement revive spontaneously during the evolution, while the third one needs a local operation to obtain the desired entanglement revival.

\subsection{Random external field}\label{sec:REF}
As mentioned in the introduction, when the local environment is a quantum non-Markovian one (for instance, a bosonic reservoir of photons inside a high-quality factor cavity \cite{bellomo2007PRL,bellomo2008PRA}), the discovery that the entanglement between two  separated noninteracting qubits can reappear during the evolution after complete vanishing has been interpreted by repeated bipartite entanglement exchanges among the quantum parts of the global system \cite{lofrancoreview,Lopez2010PRA,darrigo2012AOP}. In fact, the initial two-qubit entanglement is redistributed between the two quantum reservoirs and between a qubit and the other qubit's reservoir and, thanks to the memory effects, returns to the two qubits with a partial loss \cite{Lopez2010PRA,chiuri2012,fanchiniSciRep}. Revivals of two-qubit entanglement were successively predicted also for local non-Markovian classical environments which do not back-react and cannot share quantum excitations, such as random telegraph noises \cite{zhou2010QIP,lofranco2012PhysScripta,bordone2012} and phase noisy lasers \cite{bellomo2012noisylaser}, that poses a very simple question: in this case, where do the initial quantum correlations go when they disappear? 

In order to answer this question, the best strategy appears that of finding a simple yet paradigmatic model which allows a straightforward treatment of the phenomenon. Starting from the phase noisy laser, which is a classical field with a randomly fluctuating phase \cite{bellomo2012noisylaser,erikaJMO,cresserOptComm}, the natural simplification is that to consider a field with a random phase assuming only two possible values \cite{lofranco2012PRA,LeggioPRA}. While the first study with such a random external field employs a model where both qubits locally interact with the noise \cite{lofranco2012PRA}, here we review the simplest case where a qubit is isolated (Fig. 1(b)) \cite{LeggioPRA}, whose all-optical experimental simulation has been realized \cite{LoFrancoNatCom} and shall be discussed in Sec.~\ref{sec:ExpObserv}. 

\begin{figure}[t]
\begin{center}
{\includegraphics[width=0.45\textwidth]{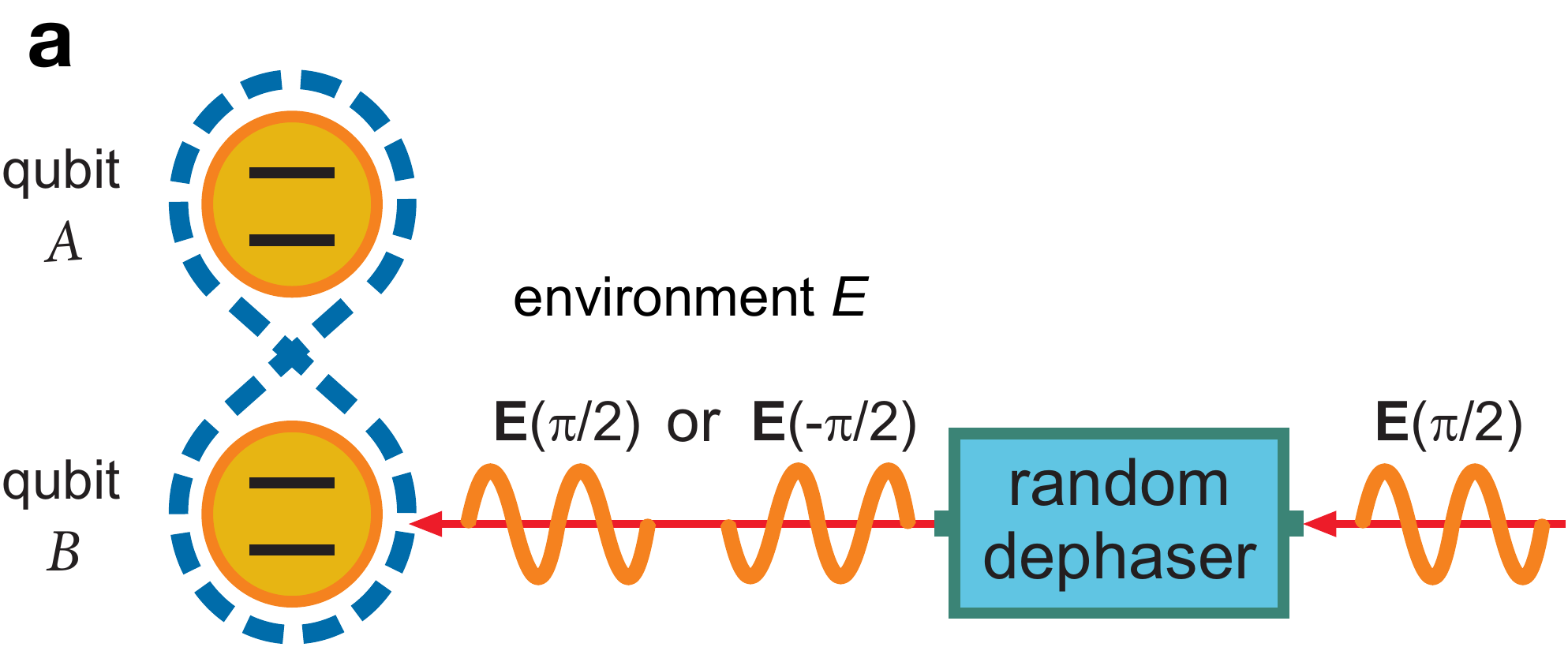}\hspace{1.5 cm}
\includegraphics[width=0.3\textwidth]{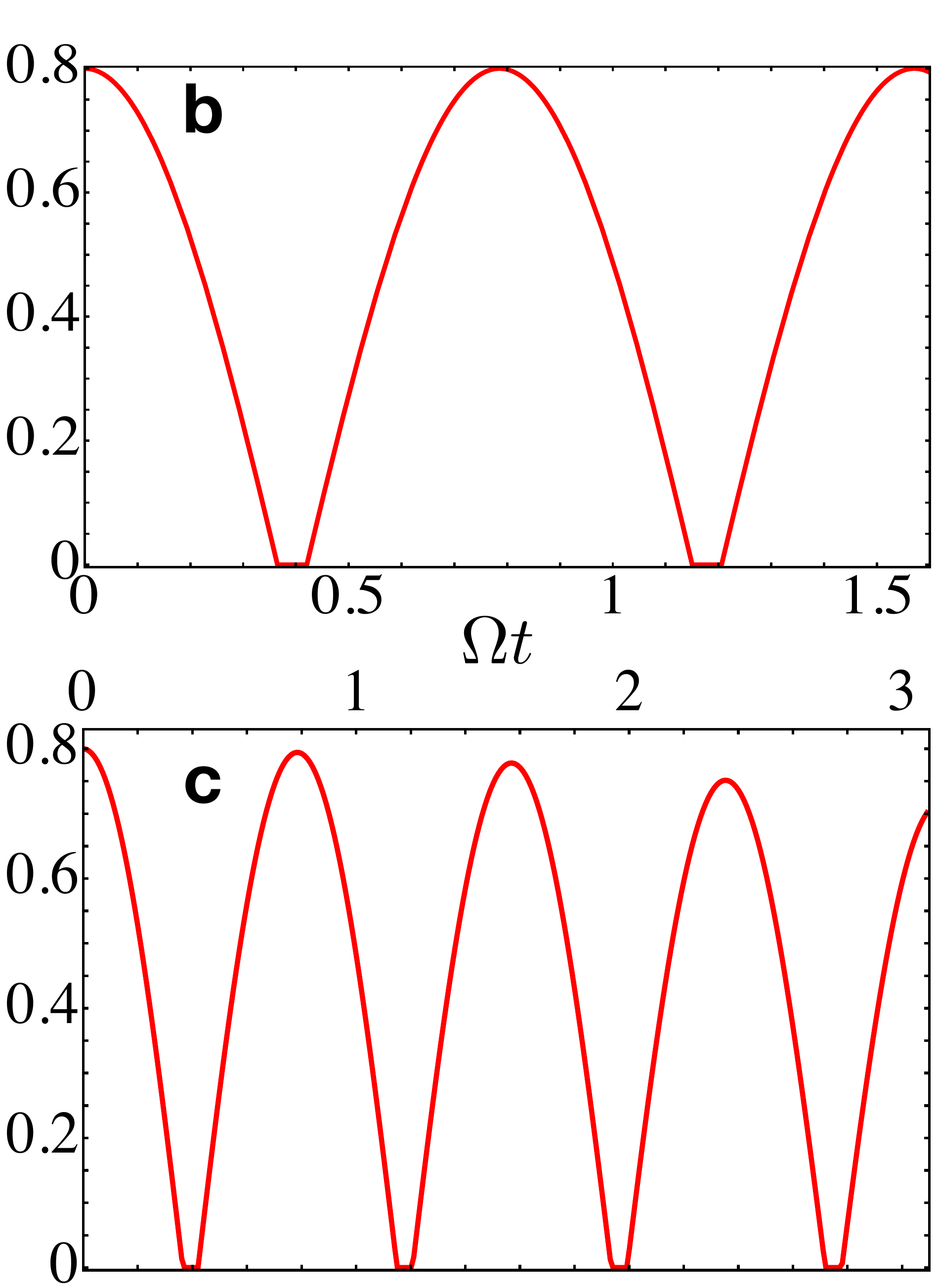}}
\end{center}
\caption{\textbf{Revivals under random external field.} \textbf{a.} A random external classical field acts on the qubit $B$. The random dephaser either shifts of $\pi$ the phase of the input field with probability $1/2$ or leaves it unchanged with probability $1/2$ (figure from Ref.~\cite{LeggioPRA}). \textbf{b.} Concurrence $C(t)$ of $\rho_{AB}(t)$ versus $\Omega t$ for initial conditions $x=z=1$ and $y=0.9$ in the case of periodic dynamics ($\sigma\rightarrow 0$). \textbf{c.} Concurrence $C(t)$ of $\rho_{AB}(t)$ for the same initial conditions in the case of decoherent dynamics ($\sigma=0.1\,\Omega$).}
\label{fig:RandomField}
\end{figure}

As illustrated in Fig.~\ref{fig:RandomField}(a), the environment is a classical field (laser) with a random phase $\varphi$, which can be $\varphi_\pm=\pm \frac{\pi}{2}$ with probability $p_{\pm}=\frac{1}{2}$ \cite{lofranco2012PRA,LeggioPRA}. The dynamical map which gives the evolved state of the two-qubit system is 
\begin{equation}\label{rhotnogauss}
\rho_{AB}^\Omega(t)=\frac{1}{2}\sum_{\varphi=\varphi_\pm}
\Big(\openone_A\otimes U_{\varphi,\Omega} (t)\Big)\rho_{AB}(0)\Big(\openone_A\otimes U_{\varphi,\Omega} ^{\dag}(t)\Big),
\end{equation}
where $\openone_A$ is the identity matrix in the Hilbert space of the qubit $A$ and  
\begin{equation}\label{unitarymatrix}
U_{\varphi,\Omega} (t)=\left(
\begin{array}{cc}\cos(\Omega t/2)&\mathrm{e}^{-\mathrm{i}\varphi}\sin(\Omega t/2)\\
-\mathrm{e}^{\mathrm{i}\varphi}\sin(\Omega t/2) & \cos(\Omega t/2) \\\end{array}\right),
\end{equation}
is the unitary matrix in the basis $\{\ket{0},\ket{1}\}$ of the time evolution operator associated to the interaction between qubit $B$ and a classical electric field $\mathbf{E}$ with phase $\varphi$. This interaction is described, in the rotating frame at the qubit-field frequency and within the rotating wave approximation, by the Hamiltonian  \cite{lofranco2012PRA,LoFrancoNatCom}
\begin{equation}\label{hamiltonian}
\mathcal{H}_\varphi=\mathrm{i}\hbar (\Omega/2)(\sigma_+e^{-\mathrm{i}\varphi}-\sigma_-e^{\mathrm{i}\varphi}),
\end{equation}
where $\Omega$ is the qubit-field coupling constant (Rabi frequency) proportional to the field amplitude and $\sigma_+=\ket{1}\bra{0}$, $\sigma_-=\ket{0}\bra{1}$ are the qubit raising, lowering operators. The non-Markovian dynamical map of Eq.~(\ref{rhotnogauss}) is a completely positive trace preserving map representing a unital channel $\Lambda_t$ (that is, $\Lambda_t \openone = \openone$) of the class of random unitaries \cite{mannone2012,erikaJMO,chruschinskirandom,alickibook}. A useful feature of this dynamical map is that, if the two-qubit initial state belongs to the class of Bell-diagonal states, which are mixtures of the four Bell states, the evolved state will remain inside this class during the evolution \cite{PhysRevA.77.042303,bellomo2012PRA2,aaronson2013PRA,aaronson2013NJP,universalfreezing}. 

In realistic situations a signal inhomogeneous broadening can be present \cite{LoFrancoNatCom} whose effect is a Gaussian distribution in the field amplitude and thus in the Rabi oscillation frequency $\Omega_g$, which must be traced out in order to get the evolved two-qubit state $\rho_{AB}(t)$. In this case one has
\begin{equation}\label{rhotgauss}
\rho_{AB}(t)=\int_{-\infty}^{\infty} d\Omega_g \, G(\Omega_g)\, \rho_{AB}^{\Omega_g}(t),\quad
G(\Omega_g)=\frac{1}{\sigma \sqrt{\pi}}e^{-\frac{(\Omega_g-\Omega)^2}{4\sigma^2}},
\end{equation} 
where $\Omega$ is the Rabi frequency without dissipation (the central Rabi frequency) and $\sigma$ the standard deviation (the Rabi frequency width). The effect of the noise on the random field is transferred to the intrinsic evolution of the quantum system. 

To investigate the dynamics originating from different initial conditions, a convenient two-qubit initial state is \cite{LeggioPRA}
\begin{equation}\label{initstate}
\rho_{AB}^0(x,y,z)=y|x_+\rangle\langle x_+|+(1-y)|z_-\rangle\langle z_-|,
\end{equation}
where
\begin{equation}
|x_+\rangle = x |2_+\rangle+\sqrt{1-x^2}|1_+\rangle,\quad
|z_-\rangle= z |2_-\rangle+\sqrt{1-z^2}|1_-\rangle,
\end{equation}
and $|1_{\pm}\rangle=(\ket{01}\pm\ket{10})/\sqrt{2}$, $|2_{\pm}\rangle=(\ket{00}\pm\ket{11})/\sqrt{2}$ are the one-excitation and two-excitation Bell (maximally entangled) states. Such an initial state allows both a linear combination (quantum coherence) between Bell states of different kinds and a statistical mixture of them. Here we limit to the case of an initial Bell diagonal state and utilize the concurrence $C$ \cite{wootters1998PRL} to quantify the two-qubit entanglement. For convenience, we recall that the concurrence is defined as $C(\rho_{AB})=\textrm{max}\{0,\sqrt{\chi_{1}}-\sqrt{\chi_{2}}-\sqrt{\chi_{3}}-\sqrt{\chi_{4}}\}$, where $\chi_{j}$'s are the eigenvalues in decreasing order of the matrix $\rho_{AB}(\sigma_{y}\otimes\sigma_{y})\rho_{AB}^{\ast}(\sigma_{y}\otimes\sigma_{y})$ with $\sigma_{y}$ denoting the second Pauli matrix and $\rho_{AB}^{\ast}$ corresponding to the complex conjugate of the two-qubit density matrix $\rho_{AB}$ in the canonical basis $\{|00\rangle,|01\rangle,|10\rangle,|11\rangle\}$.

In Fig.~\ref{fig:RandomField}(b)-(c) the dynamics of entanglement is plotted starting from an initial Bell-diagonal state $\rho_{AB}^0(1,0.9,1)$, which has a concurrence $C=0.8$ and is the same initial state considered in the experiment of Ref.~\cite{LoFrancoNatCom}. Fig.~\ref{fig:RandomField}(b) shows the periodic evolution corresponding to the case of fixed qubit-field coupling (that is, fixed Rabi frequency, $\sigma = 0$) \cite{lofranco2012PRA}; Fig.~\ref{fig:RandomField}(c) instead represents the case when the Gaussian distribution of the Rabi frequency of Eq.~\eqref{rhotgauss} is considered and entanglement peaks decay with a decoherence time proportional to $\sigma^{-1}$. The periodic dynamics can be meant as the dynamics of the system at times much shorter than the (Gaussian-induced) decoherence time. Revivals and dark periods of entanglement spontaneously shows up in both cases. The interpretations related to this model shall be discussed in Sec.~\ref{sec:Interp}.

\subsection{Local pulse under low-frequency noise}\label{sec:lowfreqnoise}

Here we briefly review the model of a pure-dephasing classical noise, that gathers basic characteristics of many nanodevices under low-frequency noise \cite{FalciPrl05,Ithier2005,bylander2011NatPhys,Chiarello}, with a local pulse applied at a certain time of the evolution \cite{darrigo2012AOP}. The Hamiltonian which rules the dynamics of the open qubit $B$ is given by ($\hbar=1$)  
\begin{equation}\label{hamiltonian-Echo}
{\cal H}_B(t)=  [-\Omega_A \sigma_{z} + \varepsilon(t) \sigma_{z} + \mathcal{V}(t) \sigma_{x}]/2,
\end{equation}
where $\varepsilon(t)$ is a stochastic process and $\mathcal{V}(t)$ an external control field. This $\mathcal{V}(t)$ represents an echo $\pi$-pulse at a given time $\overline{t}$ with evolution operator $\mathrm{e}^{-\mathrm{i}\sigma_x\pi/2}=-\mathrm{i}\sigma_{x}$, short enough to neglect the effect of noise during its application. The stochastic process has an exponential autocorrelation function $\langle \varepsilon (t) \varepsilon (0) \rangle=\sigma^2 \mathrm{e}^{-t/\tau}$,
with noise correlation time $\tau$.
For simplicity, the stochastic process $\varepsilon(t)$ is chosen slow enough to be approximatively static $\varepsilon(t)\approx \varepsilon$ during the evolution time $t$, which means $\tau\to\infty$. The parameter $\varepsilon$ is a Gaussian random variable with zero expectation value and standard deviation $\sigma$. This static noise produces an effect analogous to inhomogeneous broadening in nuclear magnetic resonance (NMR) \cite{vandersypen2005RMP}. 

\begin{figure}[t]
\begin{center}
\includegraphics[width=0.45\textwidth]{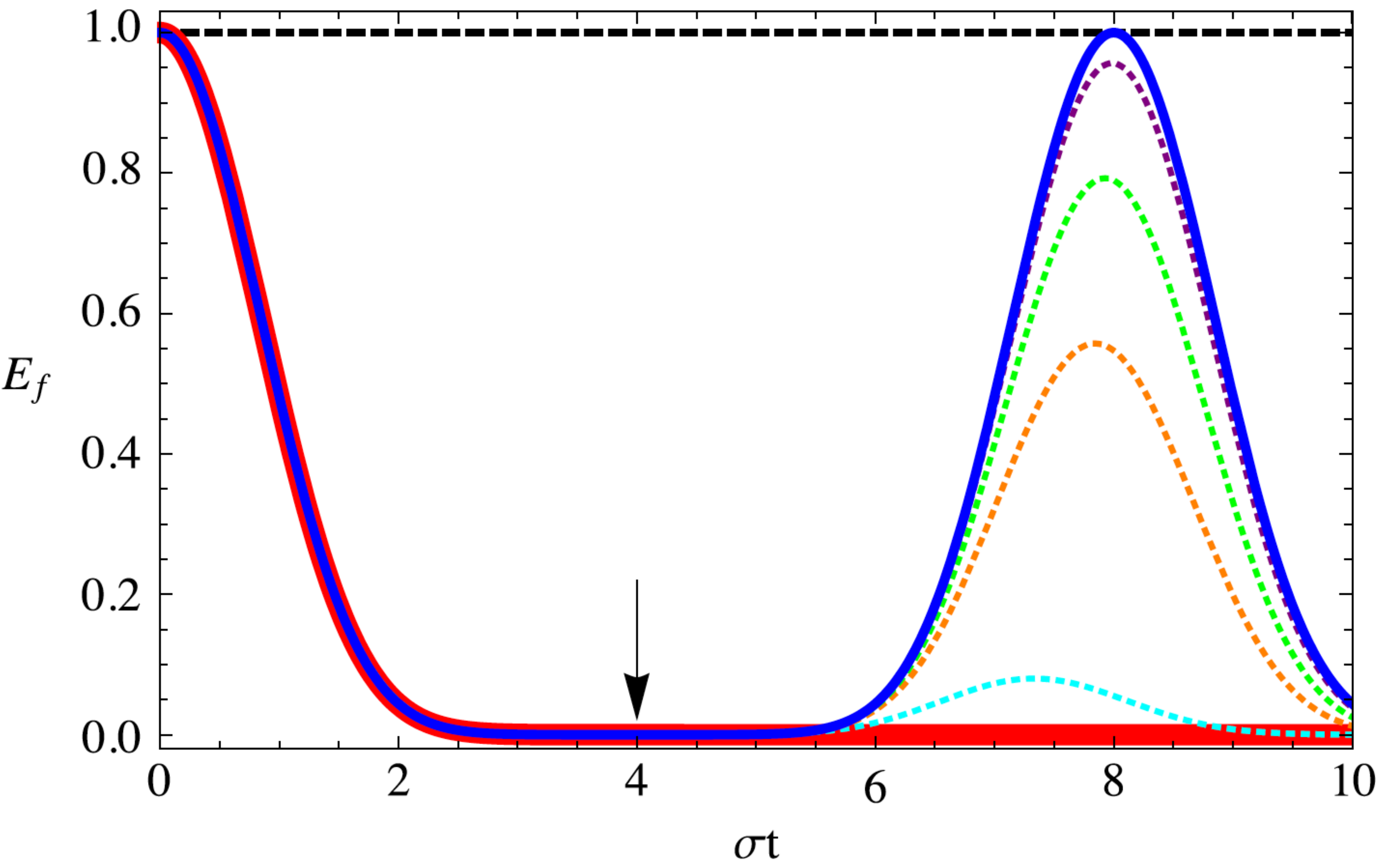}
\end{center}
\caption{\textbf{Revival by local operation under low-frequency noise.} Entanglement of formation $E_f(\rho(t))$ as a function of the dimensionless time $\sigma t$. 
The thick red line gives the free evolution under static noise while the thin blue solid line represents the evolution when an echo pulse is applied at time $\sigma\overline{t}=4$ (individuated by the black arrow). The black dashed line is the average entanglement of the system $E_\mathrm{av}=1$. Dotted lines represent $E_f(\rho(t))$ for a non-static $\varepsilon(t)$ with increasing values of $\sigma\tau$ from bottom to top. Total entanglement recovery is obtained only in the limit of static noise ($\tau/\overline{t}\to\infty$). Figure from Ref.~\cite{darrigo2012AOP}.}
\label{fig:LocalOperation}
\end{figure}

Taking the two qubits initially in any of the four Bell states, indicated with $\ket{\Psi_0}$, applying the evolution operator due to the Hamiltonian of Eq.~(\ref{hamiltonian-Echo}) and tracing out the static noise degrees of freedom, one finds that the two-qubit system evolves in a mixed state $\rho(t)=\int d\varepsilon p(\varepsilon) |\Psi_\varepsilon(t)\rangle\langle \Psi_\varepsilon(t)|$, where $\ket{\Psi_\varepsilon(t)}=\hat{T}\mathrm{e}^{-\mathrm{i}\int_0^t{{\cal H}_A(t')dt'}}\otimes \hat{T}\mathrm{e}^{-\mathrm{i}\int_0^t{{\cal H}_B(t')dt'}} \ket{\Psi_0}$ and $p(\varepsilon)$ is the Gaussian probability density function of $\varepsilon$. The corresponding time-dependent concurrence is given by \cite{darrigo2012AOP}
\begin{equation}\label{twoqubitconc}
C(\rho(t))=\left\{
\begin{array}{ll}
\mathrm{e}^{-\frac{1}{2}\sigma^2 t^2},& 0\leq t\leq\overline{t},\\
\mathrm{e}^{-\frac{1}{2}\sigma^2 (t-2\overline{t})^2},& \overline{t}< t\leq 2\overline{t}.
\end{array}
\right .
\end{equation}
In Fig.~\ref{fig:LocalOperation} the entanglement of formation $E_f(\rho(t))$ is plotted, which is monotonically related to the concurrence by \cite{wootters1998PRL}
\begin{equation}\label{eq:wootters-formula}
E_f(\rho(t))=\mathtt{h}\Big(\frac{1+\sqrt{1-C(\rho(t))^2}}{2}\Big), 
\end{equation}
where $\mathtt{h}(x)=-x\log_2x-(1-x)\log_2(1-x)$.
It is displayed that, if no pulse is applied, $E_f(\rho(t))$ decays and tends to zero at times $\sigma t\gg 1$.  
Differently, the action of a local pulse at $t=\overline{t}$ makes $E_f(\rho(t))$ revive and reach its initial maximum value $E_f(\rho(2\overline{t}))=E_f^\textrm{max}=1$. This value coincides with the average entanglement $E_\mathrm{av}({\cal A}(t))=\int p(\varepsilon) E_f(\ket{\Psi_\varepsilon(t)}\bra{\Psi_\varepsilon(t)}) d\varepsilon=1$ of the evolved physical ensemble ${\cal A}=\{p(\varepsilon)d\varepsilon,\,\ket{\Psi_\varepsilon(t)}\}$ \cite{darrigo2012AOP}. Notice that in this situation the entanglement revival is not spontaneously found during the evolution but created by means of the local pulse, which makes the dynamical map indivisibile and thus non-Markovian \cite{darrigo2012AOP,rivas2010PRL,guo-piilo,addisNJP}.
A discussion about the interpretation of this result shall be reported in Sec.~\ref{sec:Interp}.

\subsection{Random telegraph noise}

The system we review here consists in a pair of independent superconducting qubits, $A$ and $B$, where qubit $B$ interacts with a bistable impurity (fluctuating charge) which produces pure dephasing RTN \cite{lofranco2012PhysScripta}. The Hamiltonian of qubit $B$ is ($\hbar=1$) \cite{paladino2002PRL}
\begin{equation}
\mathcal{H}_B=-(\omega_0/2)\sigma_z-(v/2)\xi(t)\sigma_z,
\end{equation} 
where $\xi(t)$ establishes the RTN switching at a rate $\gamma$ between $\pm1$ and $v$ is the qubit-RTN coupling constant. The ratio $g=v/\gamma$ is the characteristic parameter that rules the crossover between a Markovian noise for weakly coupled impurities ($g<1$) and a non-Markovian noise for strong coupled impurities ($g>1$) \cite{paladino2002PRL}. The exact evolution of single-qubit coherence $q(t)\equiv\rho_{01}(t)/\rho_{01}(0)$ is known \cite{lofranco2012PhysScripta,paladino2002PRL} and in turn allows to obtain the evolved two-qubit density matrix by a standard procedure based on the independence of the two qubits and their own environments \cite{bellomo2007PRL,bellomo2008PRA}. 

With the qubits initially prepared in the extended Werner-like (EWL) states \cite{bellomo2008PRA}
\begin{equation}\label{EWLstates}
    \rho_1=r \ket{1_{a}}\bra{1_{a}}+\frac{1-r}{4}\openone_4,\quad
    \rho_2=r \ket{2_{a}}\bra{2_{a}}+\frac{1-r}{4}\openone_4,
\end{equation}
where $\ket{1_{a}}=a\ket{01}+b\ket{10}$, $\ket{2_{a}}=a\ket{00}+b\ket{11}$ with $|a|^2+|b|^2=1$ and $\openone_4$ is the two-qubit identity matrix, one can follow the entanglement dynamics by the concurrence $C=C(t)$. The density matrix of EWL states has an X form \cite{bellomo2008PRA} and this structure is maintained during the pure dephasing evolution. We recall that entangled states of superconducting qubits with purity $\approx 0.87$ and fidelity to Bell states $\approx 0.90$ have been experimentally generated \cite{schoelkopf2009Nature} and can be approximately described by EWL states with a purity parameter $r_\mathrm{exp}\approx0.91$.

\begin{figure}[t]
\begin{center}
\includegraphics[width=0.45\textwidth]{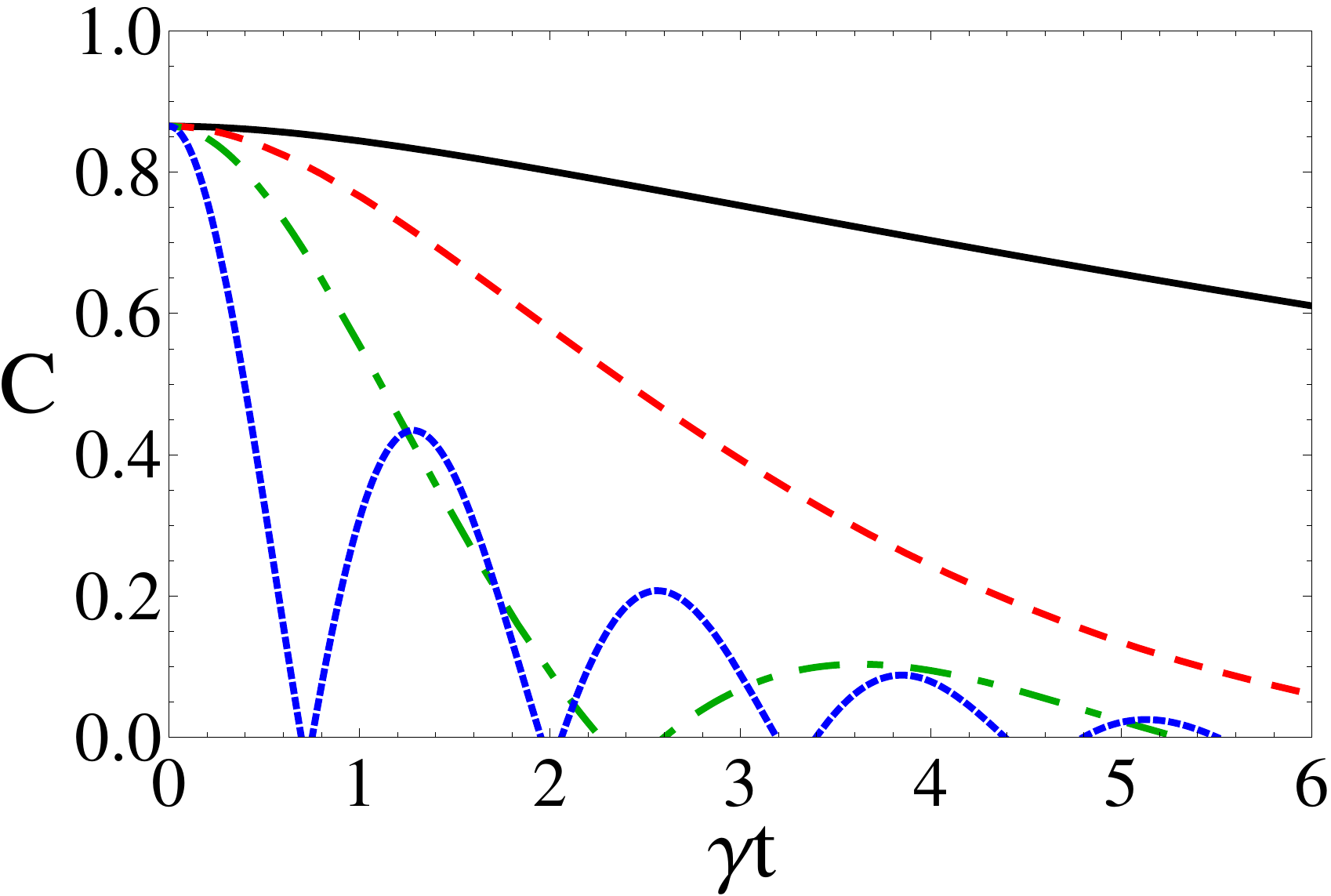}
\end{center}
\caption{\textbf{Revivals under random telegraph noise.} Concurrence as a function of the dimensionless time $\gamma t$ for values of $g$ equal to $0.5$ (solid black line), $1.1$ (red dashed line), $2$ (green dot-dashed line), $5$ (blue dotted line). Initial state parameters are $r=r_\mathrm{exp}=0.91$, $a=b=1/\sqrt{2}$. Figure from Ref.~\cite{lofranco2012PhysScripta}.}
\label{fig:RTN}
\end{figure}

The concurrences at time $t$ for both the two initial states of Eq.~(\ref{EWLstates}) are equal to $C(t)=\mathrm{max}\{0,2K(t)\}$, where $K(t)=r|a|\sqrt{1-|a|^2}|q(t)|-(1-r)/4$, $q(t)$ being the single-qubit coherence. 
The plots of Fig.~\ref{fig:RTN} display that a sequence of revivals of entanglement occur provided that the coupling parameter $g$ reaches sufficiently high values which enable a non-Markovian dynamics for the system, with the frequency of revivals increasing as $g$ increases. Therefore, a noise of completely classical nature as the RTN causes two-qubit entanglement to reappear after dark periods once the non-Markovian features of the noise has been activated by a sufficiently strong qubit-impurity coupling \cite{zhou2010QIP,lofranco2012PhysScripta,bordone2012}.  

\section{Experimental observations}\label{sec:ExpObserv}

This section deals with the description of two all-optical experiments that, simulating two of the models reported above, confirm that quantum entanglement can either spontaneously revive \cite{LoFrancoNatCom} or be recovered by local operations \cite{adeline2014} in classical environments.   

\subsection{Experimental entanglement revivals under random external field}\label{sec:Exprevival}
\begin{figure}[t]
\begin{center}
\includegraphics[width=0.52\textwidth]{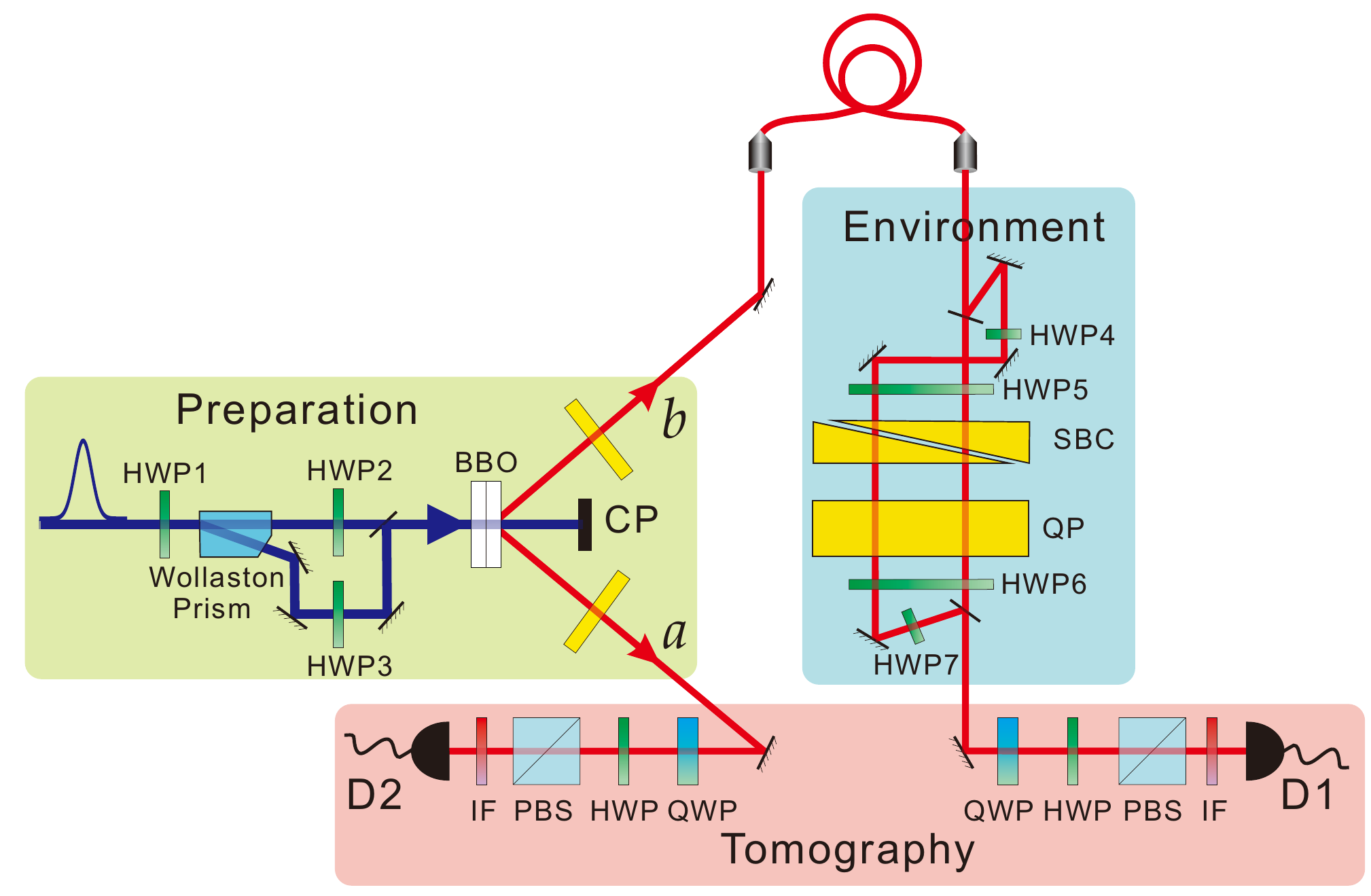}
\end{center}
\caption{\textbf{Experimental setup simulating the random external field.} The all-optical setup which realizes the model of Fig.~\ref{fig:RandomField}(a) is made of three main parts. (i) \textit{Preparation}. This part is devoted to initialize the two-photon state in the desired Bell-diagonal state. The photon in mode $a$ (corresponding to the isolated qubit $A$) is directly sent to the measurement apparatus. (ii) \textit{Environment}. The photon in mode $b$ (representing the open qubit $B$) reaches the environment part, with two possible probabilistic paths representing the two unitaries (random phases). (iii) \textit{Tomography}. This part performs suitable measures of the photon polarizations which allow the  construction of the output (evolved) density matrix. Figure from Ref.~\cite{LoFrancoNatCom}.}
\label{fig:setupHefei}
\end{figure}

The simplicity of the theoretical model of Fig.~\ref{fig:RandomField}(a) has the advantage of making it realizable by a neat experimental setup which avoids any side effect that can influence the expected dynamics and complicate its interpretation.
In particular, an all-optical experiment was reported that simulates this model, with the random external field mimicked by quantum degrees of freedom of the optical devices, and allows observation and control of entanglement revivals without system-environment back-action \cite{LoFrancoNatCom}. The experimental setup is shown in Fig.~\ref{fig:setupHefei}. The bipartite quantum system is made of two polarized photons, each one representing a qubit with basis states $\ket{H}$ (horizontal polarization) and $\ket{V}$ (vertical polarization). We omit the very technical details of the devices employed in the setup (available in Ref.~\cite{LoFrancoNatCom}) while focusing on the general aspects of the experiment which determine the realization of the target model. 

The \textit{preparation part} of the setup generates a pair of polarization entangled photons in a desired Bell-diagonal state $\rho_{ab}^{\mathrm{in}}$. The photon in mode $a$ (the isolated qubit) is directly sent to the state tomography part while the photon in mode $b$ goes to the environment part and finally to state tomography part.

The \textit{environment part} of the setup simulates the random external field on qubit $b$ by exploiting a beam-splitter that creates two photon paths (reflected $\mathrm{p}_+$ and transmitted $\mathrm{p}_-$), corresponding to the effect of the field with either phase, plus the measurement process that does not distinguish the two paths $\mathrm{p_\pm}$ in a classically probabilistic fashion, creating a statistical mixture of them with equal probabilities ($1/2$). 
The two photonic paths are designed such as to induce, apart from an unimportant global phase factor, the unitary transformations\begin{eqnarray}\label{HVtransform}
\ket{H}&\stackrel{\mathrm{p}_\pm}{\longrightarrow}&\cos(\phi/2)\ket{H}\pm i\sin(\phi/2)\ket{V},\nonumber\\
\ket{V}&\stackrel{\mathrm{p}_\pm}{\longrightarrow}&\pm i\sin(\phi/2)\ket{H}+\cos(\phi/2)\ket{V},
\end{eqnarray}
where $\phi$ is the phase difference between $\ket{H}$ and $\ket{V}$ introduced by the Soleil-Babinet compensator (SBC) and the quartz plates (QPs). This phase difference is defined as $\phi=\omega\tau$, where $\omega$ is the photon frequency and $\tau\equiv L\Delta n/c$ is the time taken by the photon to cross the optical element (SBC or QP), $L$ being the thickness of the optical element, $c$ the vacuum speed of light, $\Delta n$ the difference between the refraction indices of $H$ and $V$ polarizations. 
It is immediate to see that the two paths $\mathrm{p}_\pm$ of Eq.~(\ref{HVtransform}) define unitaries $U_{\mathrm{p}_\pm}(\phi)$ on the basis states $\{\ket{H},\ket{V}\}$ of $b$ which act exactly as the two time evolution operators $U_{\varphi_\mp}(t)$ of Eq.~(\ref{unitarymatrix}) on the qubit $B$, respectively, with the connections $\ket{0}\leftrightarrow\ket{H}$, $\ket{1}\leftrightarrow\ket{V}$ and $\phi=\omega\tau \leftrightarrow \Omega t$. The overall output state thus becomes
\begin{equation}
\Lambda\rho_{ab}^{\mathrm{in}}=\frac{1}{2}\sum_{p=\mathrm{p}_\pm}(\openone_a\otimes U_{p})\rho_{ab}^{\mathrm{in}} (\openone_a\otimes U_{p}^\dag), 
\end{equation}
which reproduces the two-qubit evolved density matrix of Eq.~(\ref{rhotnogauss}). In the experiment the photon has an intrinsic Gaussian frequency distribution $f(\omega)=(2/\sigma\sqrt{\pi})\exp[-4(\omega-\omega_0)^2/\sigma^2]$ \cite{xu2009PRL,xu2010NatComm}, where $\omega_0$ is the center frequency and $\sigma$ the frequency width (standard deviation). This frequency degree of freedom is a decoherence source analogous to that due to the Rabi frequency distribution in the model of Fig.~\ref{fig:RandomField}(a) described above. The experimental evolved state $\rho_{ab}^{\mathrm{out}}$ is then determined by tracing out the photon frequency stochastic variable from $\Lambda\rho_{ab}^{\mathrm{in}}$, giving rise to an evolved state analogous to that of Eq.~(\ref{rhotgauss}). 

The \textit{tomography part} finally performs standard quantum state tomography for constructing the output (evolved) density matrix $\rho_{ab}^{\mathrm{out}}$ of the two photons at many values of the experimental time $\tau$.

\begin{figure}[t]
\begin{center}
\includegraphics[width=0.75\textwidth]{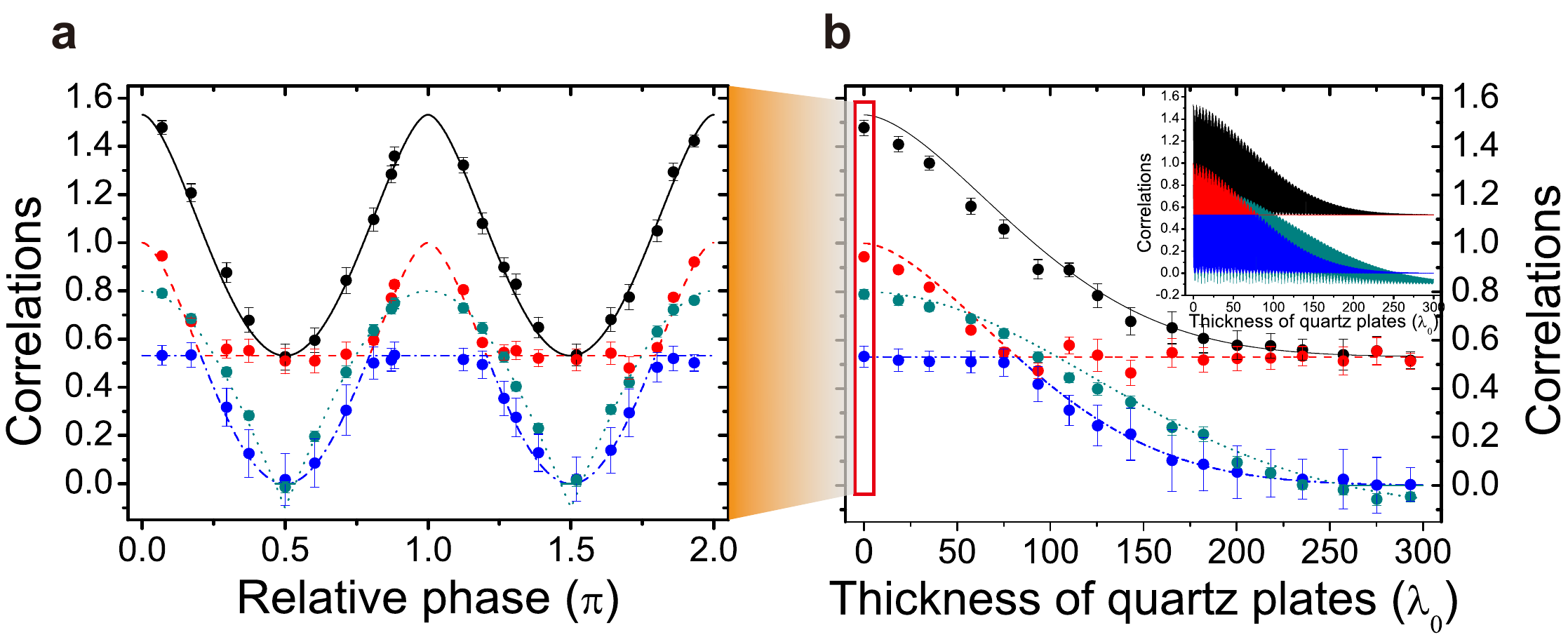}
\end{center}
\caption{\textbf{Experimental observations for the random external field.} \textbf{a.} Theoretical (cyan line) and experimental (cyan points) two-qubit entanglement as a function of the relative phase in the coherent evolution. Only the SBC is used in the setup and the relative phase (in units of $\pi$) is $\phi=\omega_0\tau$. Revivals of entanglement are clearly visible. \textbf{b.} Theoretical (cyan line) and experimental (cyan points) envelope of entanglement dynamics as a function of the quartz plate length $L$ (in units of $\lambda_0=800$nm, that is the center photon wavelength). The coherent evolution with revivals in panel \textbf{a} is a part of the decoherent evolution, evidenced by the red box in panel \textbf{b}. The inset shows the theoretical dynamics of the various correlations. In both panels, the black, red and blue lines (points) represent, respectively, the theoretical (experimental) total correlations, classical correlations and quantum discord, whose description is out of the scopes of this review. The initial state $\rho_{ab}^{\mathrm{in}}$ is the Bell-diagonal state $\rho_{AB}^0(1,0.9,1)$ of Eq.~(\ref{initstate}). Figure from Ref.~\cite{LoFrancoNatCom}.}
\label{fig:DynamicsHefei}
\end{figure}

The two-photon system is initialized in the Bell-diagonal state $\rho_{ab}^{\mathrm{in}}=\rho_{AB}^0(1,0.9,1)$ of Eq.~(\ref{initstate}). The entanglement evolution is followed by resorting to the concurrence $C(\rho_{ab}^{\mathrm{out}})$, the experimental points being acquired from the reconstructed output density matrices by state tomography. 
The results for the coherent evolution, where only the SBC is used, are plotted as a function of the relative phase $\phi=\omega_0\tau$ in Fig.~\ref{fig:DynamicsHefei}(a). Entanglement exhibits dark periods (around points $\pi/2$ and $3\pi/2$ in a $2\pi$ period) and revivals.
In Fig.~\ref{fig:DynamicsHefei}(b) the experimental results are displayed for the decoherent evolution, where both SBC and QPs are used. In particular, the main panel shows the envelope dynamics of correlations as a function of the quartz plate length $L$, given by the maximum amplitudes of revivals. The maxima of entanglement revivals monotonously decrease and totally vanishes at $L\approx 258\lambda_{0}$. The inset of Fig.~\ref{fig:DynamicsHefei}(b) is a plot of the theoretical curves exhibiting these decaying revivals. The coherent evolution is part of the decoherent evolution, as highlighted by the red box in Fig.~\ref{fig:DynamicsHefei}(b). 

\subsection{Experimental entanglement revival by local operations under low-frequency noise}
An all-optical experiment \cite{adeline2014} has been reported which reproduces the theoretical model of a longitudinal low-frequency noise with a local echo pulse, described above and ruled by the Hamiltonian of Eq.~(\ref{hamiltonian-Echo}). We review here the main aspects and results of this experiment.

\begin{figure}[t]
\begin{center}
\includegraphics[width=0.76\textwidth]{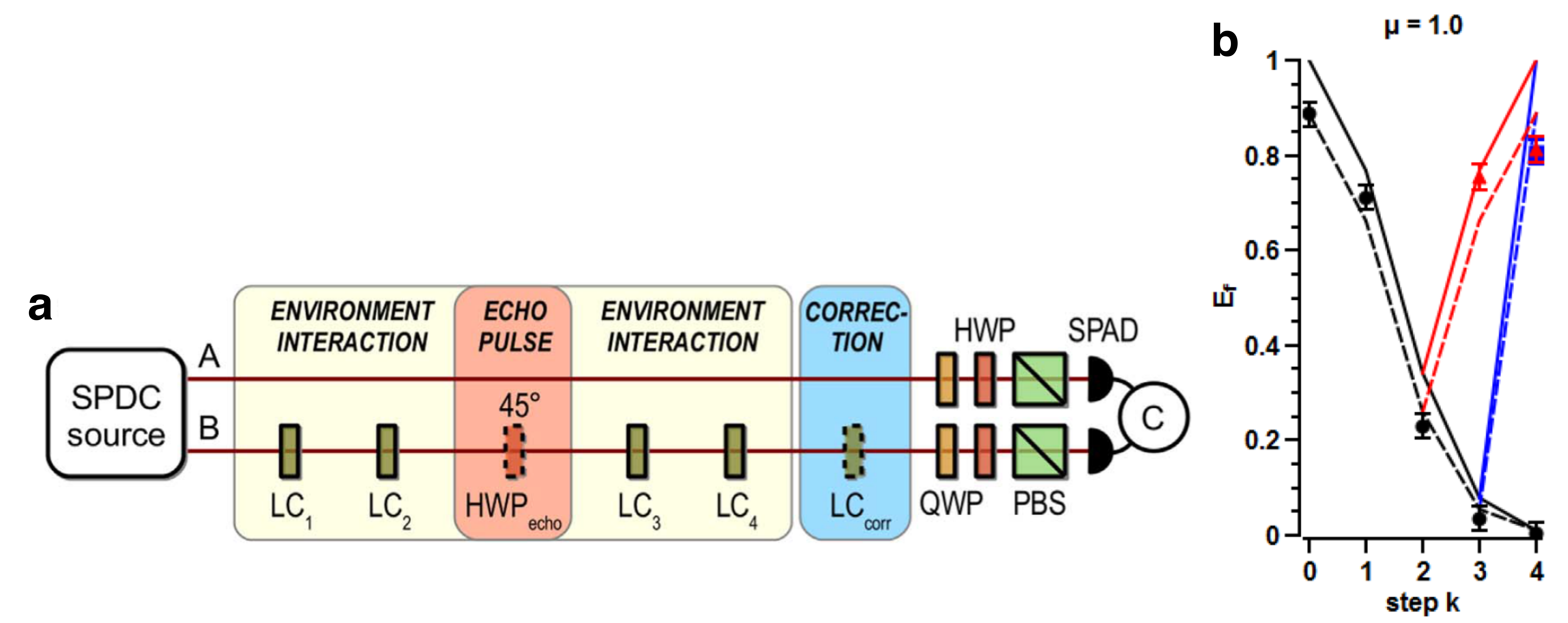}
\end{center}
\caption{\textbf{Experimental setup and observations for the low-frequency noise with local operation.} \textbf{a.} The experimental apparatus initially prepares qubits $A$ and $B$ in the Bell state $\ket{\Psi^-}$ by standard parametric down conversion (SPDC). While qubit $A$ directly goes to measurement part, qubit $B$ stroboscopically interacts with the environment through four random phases induced by liquid crystal (LC) retarders. The noise induced by the environment is compensated by an echo-pulse unitary $U_\mathrm{echo}=\sigma_x$ produced by an half-wave plate (HWP). Other elements in the measurement part are quarter-wave plate (QWP), polarizing beam-splitter (PBS), single photon avalanche photodiode (SPAD) and coincidence counting electronics (C). The ``correction'' $\mathrm{LC}_\mathrm{corr}$ represents a rephasing unitary which is able to compensate the dephasing noise when the latter is known (a situation that is not treated here). \textbf{b.} Entanglement of formation $E_f$ measured at each step $k$ ($k=1,2,3,4$) for $\mu=1$ (static noise). Points and lines represent the experimental data 
and the theoretical calculations, respectively. Dashed lines are simulations for a state with a fidelity $F= 0.96$ to a Bell state. Black and red colors correspond, respectively, to the uncontrolled and pulsed dynamics (the blue one is the controlled dynamics). Figures from Ref.~\cite{adeline2014}.}
\label{fig:setupRoma}
\end{figure}

The experimental setup is illustrated in Fig.~\ref{fig:setupRoma}(a). The qubit information is encoded, as usual, in the horizontal $H$ and vertical $V$ polarizations of two photons $A$ and $B$ propagating in the freespace. The two-photon system is initially prepared in the (maximally entangled) Bell state $\ket{\Psi^-}=(\ket{H_AV_B}-\ket{V_AH_B})/\sqrt{2}$ by standard parametric down-conversion. Photon $A$, representing the isolated qubit, is directly sent to the measurement device, whilst photon $B$ interacts with a classical environment described by a stochastic process $x(t)$, playing the role of the variable $\epsilon(t)$ in the Hamiltonian of Eq.~(\ref{hamiltonian-Echo}). 

The designed noisy channel acting on $B$ induces pure dephasing at times $t_k$ by means of a sequence of four liquid crystals retarders (LC$_k$), each one introducing a phase $x_k \equiv x(t_k)$ between the photon polarization components, that is 
$\alpha\ket{H_B}+\beta\ket{V_B} \rightarrow \alpha\ket{H_B}+e^{i x_k}\beta\ket{V_B}$. This procedure realizes the desired interaction Hamiltonian $H_B(t)=x(t)\delta(t-t_k)\sigma_z/2$, where $\sigma_z = \ket{H}\bra{H}-\ket{V}\bra{V}$ and $\delta(t)$ is the Dirac delta function. The induced phase $x_k\in[0,\pi]$ can be arbitrarily adjusted by the voltage applied to each LC$_k$. The stochastic process is simulated by generating an ensemble of $N$ random phase sequences $\{x_1,x_2,x_3,x_4\}$, where each phase $x_k$ is a Gaussian random variable with same variance $\sigma^2$ and (normalized) autocorrelation $\mu\equiv \ave{x_k x_{k+1}}/\sigma^2$ ($\mu\in[0,1]$). The local echo pulse on photon $B$ is then produced by means of a half-wave plate (HW) at $45^\circ$ between LC$_2$ and LC$_3$ (see Fig.~\ref{fig:setupRoma}(a)) which realizes a local bit-flip operation $U_\mathrm{echo}=\sigma_x$, flipping the polarization of photon $B$ ($\sigma_x\ket{H}=\ket{V}$ and viceversa).   
The dynamics of the two-photon system, which must be averaged with respect to all the phase sequences in order to trace out the noise degrees of freedom, is finally determined by mixing together the tomographic measurement data obtained for each realization of the $N$ random phase sequences \cite{adeline2014}. Here, we focus on the case of static noise which is produced for $\mu=1$ (see discussion after Eq.~(\ref{hamiltonian-Echo})). 

The evolution of the entanglement of formation of the two photons is shown in Fig.~\ref{fig:setupRoma}(b). In absence of local control, entanglement monotonously decays as evidenced by black points and lines. Differently, entanglement is recovered when a local pulse is applied, as displayed by red points and lines. An entanglement echo is thus realized in the system dynamics as predicted by the theoretical model \cite{darrigo2012AOP}.

\section{Interpretations of the phenomenon}\label{sec:Interp}

So far, three interpretations for the phenomenon of revivals of entanglement in classical environments have been proposed that can be summarized as follows: 

\begin{itemize}
\item[(i)] the classical environment plays a role as a control mechanism which keeps a classical record for what operation has been applied to the quantum system \cite{lofranco2012PRA,LoFrancoNatCom};
\item[(ii)] there is an interchange between threepartite correlations and two-qubit entanglement due to system-environment information flows \cite{LeggioPRA}; 
\item[(iii)] the quantum system contains hidden entanglement \cite{darrigo2012AOP}, that is the amount of quantum correlations not revealed by the density matrix description of the system which is recoverable by local operations \cite{darrigo2012AOP,adeline2014,lofrancoPRB}.
\end{itemize}

In this section we review these interpretations, providing the physical aspect that put them under a unified point of view.

\subsection{Quantum-classical state and classical environment as a controller}\label{sec:classcontroller}

The model with random external field described in Sec.~\ref{sec:TheorPredictions} can be conveniently described by means of a quantum-classical state, the quantum part played by the two qubits $A$, $B$ and the classical part by the environment $E$\cite{lofranco2012PRA,LoFrancoNatCom}. For the sake of clearness, we consider here the case when decoherence due to the Gaussian distribution of the Rabi frequency is negligible (analogous argumentations hold even if decoherence is present \cite{LeggioPRA}). Since the classical environment can only be in a time-invariant maximal mixture of its basis states, the overall initial state can be written as
\begin{equation}
\rho_{ABE}(0)=\rho_{AB}(0)\otimes\rho_{E}=\rho_{AB}(0)\otimes\frac{1}{2}\sum_{\varphi=\varphi_\pm}|\varphi\rangle\langle \varphi|, 
\end{equation}
where $|\varphi_+\rangle$ ($|\varphi_-\rangle$) corresponds to the state of the field with phase $\varphi=\frac{\pi}{2}$ ($\varphi=-\frac{\pi}{2}$). 
It is then possible to define a unitary evolution $U_{BE}(t)$ acting on the bipartition $B$-$E$
\begin{equation}\label{unitaryBE}
U_{BE}(t)=\sum_{\varphi=\varphi_\pm} U_{\varphi,\Omega} (t) \otimes  |\varphi\rangle\langle \varphi|,
\end{equation}
where $U_{\varphi,\Omega} (t)$ is the unitary operator of Eq.~(\ref{unitarymatrix}). By the introduction of $U_{BE}(t)$, the evolved state of the threepartite system $ABE$ is obtained by
\begin{equation}\label{overallstatet}
\rho_{ABE}(t)=[\openone_A\otimes U_{BE}(t)]\rho_{ABE}(0)[\openone_A\otimes U_{BE}^{\dag}(t)].
\end{equation}
The two-qubit evolved state $\rho_{AB}(t)$ of Eq.~(\ref{rhotgauss}) is then straightforwardly determined by tracing out the environmental degrees of freedom ($|\varphi_+\rangle$, $|\varphi_-\rangle$ in this case) from $\rho_{ABE}(t)$. The dynamics of the open system is non-Markovian \cite{mannone2012,LoFrancoNatCom}, as witnessed by well-known measures of non-Markovianity \cite{breuer2009PRL,rivas2010PRL}. 
During the evolution due to $\openone_A\otimes U_{BE}$, the states of the classical environment remain invariant, the qubit $B$ does not influence the environment $E$ and the qubit-environment back-action is thus absent \cite{lofranco2012PRA,LoFrancoNatCom}. Moreover, a classical environment cannot store any quantum correlations on its own. The bipartition $B$-$E$ evolves under the local unitary operation $U_{BE}$ so that the quantum correlations between $B$-$E$, including entanglement, are invariant. If one traces out the isolated qubit $A$ from $\rho_{ABE}(t)$, it is easy to see that the qubit $B$ and its environment $E$ never become quantum correlated. For instance, for an initial $A$-$B$ Bell-diagonal state, like that considered in the model and in the experiment described above, the reduced state of $B$-$E$ during the evolution is the uncorrelated state $(\openone_B/2)\otimes(\openone_E/2)$. Therefore, the qubit-environment correlations do not enter the phenomenon of entanglement revivals.

\begin{figure}[t]
\begin{center}
{\includegraphics[width=0.7\textwidth]{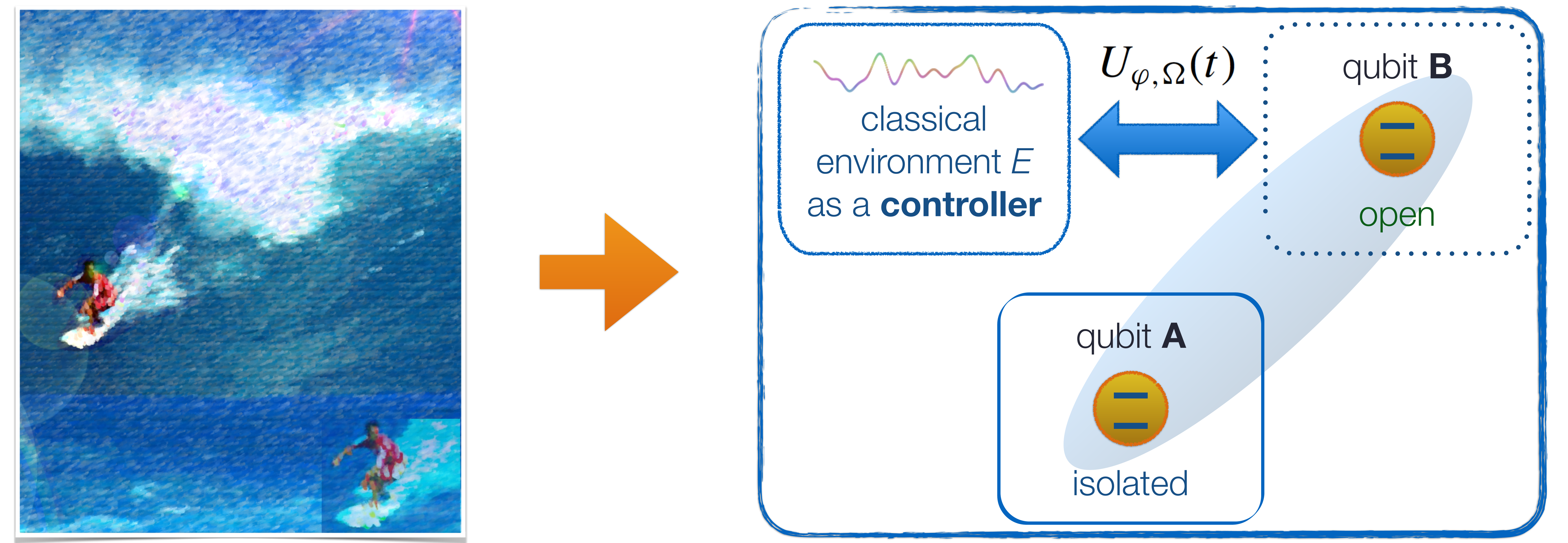}}
\end{center}
\caption{\textbf{Classical environment as a control system.} On the left, a wave rules the dynamics of two coordinated surfers where only one of them is on the wave which remains unaffected by the surfer's motion. This ``classical world'' situation may supply a pictorial description of a classical environment without back-action whose states control which unitary $U_{\varphi,\Omega} (t)$ is applied to its qubit thus determining the dynamics of the two initially correlated qubits.}
\label{fig:ClassControl}
\end{figure}

The introduction of the unitary evolution $U_{BE}(t)$ of Eq.~\eqref{unitaryBE} has a crucial role in suggesting an interpretation of these revivals by means of the role of the classical environment as a \textit{controller} for which unitary operation is acting on the system, as pictorially shown in Fig.~\ref{fig:ClassControl}. By memory effects, being the dynamics non-Markovian, the environment $E$ keeps a classical record of what unitary operation has been applied to the qubit $B$ and this occurs even without back-action. The information about the quantum system held by the environment $E$ is therefore due to what action $E$ performs on the system itself. At times when the environment loses this classical information (statistical mixing of the two different unitary operations $U_{\varphi_\pm,\Omega}(t)$, e.g., at $\Omega t=\pi/2$), entanglement disappears; at times when this information is recovered (both unitaries $U_{\varphi_\pm,\Omega}(t)$ act as the same operation, e.g., at $\Omega t=\pi$ they are equal to $\sigma_x$), entanglement revives \cite{lofranco2012PRA,LoFrancoNatCom}. 

\subsection{Tripartite correlations and information flows}

The general model of Fig.~\ref{fig:system} can be viewed from the standard decoherence paradigm of a quantum system (qubit $A$) entangled with a measurement apparatus (qubit $B$) which interacts with an environment ($E$) \cite{zurekreview}, and studying the information fluxes between the system $A$ and the environment $E$ \cite{walbornPRL,walbornPRA}. This fact allows the investigation of the mechanisms underlying the revivals of two-qubit entanglement by approaching the problem from an information-theoretic point of view. For the case of random external field of Fig.~\ref{fig:RandomField}(a), by suitably tracing out the degrees of freedom of the unwanted subsystem in the evolved threepartite state $\rho_{ABE}(t)$ of Eq.~(\ref{overallstatet}), it is straightforward to obtain the evolved reduced density matrices of the various components of the global system. Relations between the two-qubit entanglement and the genuine threepartite correlations of the system can be thus found, together with the flows of information among the different parties which provide physical grounds of this relationship \cite{LeggioPRA}.

A suitable measure of genuine tripartite correlations of the system $\{A,B,E\}$ is \cite{BellomoPRL,mazieroPRA}
\begin{equation}\label{tripartite}
\tau(\rho_{ABE})=\min\big\{I(\rho_{AB,E}),I(\rho_{AE,B}),I(\rho_{BE,A})\big\},
\end{equation}
where $I(\rho_{ij,k})=S(\rho_{ij})+S(\rho_k)-S(\rho_{ijk})$ is the quantum mutual information across any possible bipartition $ij$-$k$ of the tripartite system $\{A,B,E\}$ and $S(\rho)=-\mathrm{Tr}\rho\ln\rho$ is the von Neumann entropy of the quantum state $\rho$. We recall that genuine tripartite correlations are those which cannot be described as bipartite correlations within any bipartition of a threepartite system  \cite{tripartiteconditions}. The above measure $\tau$ takes into account both classical and quantum correlations of the hybrid quantum-classical system. To evidence dynamical relations between two-qubit entanglement and tripartite correlations, the entanglement is quantified, as usual, by the concurrence $C(\rho_{AB})$ of the two-qubit reduced state.

\begin{figure}[t]
\begin{center}
{\includegraphics[width=0.35\textwidth]{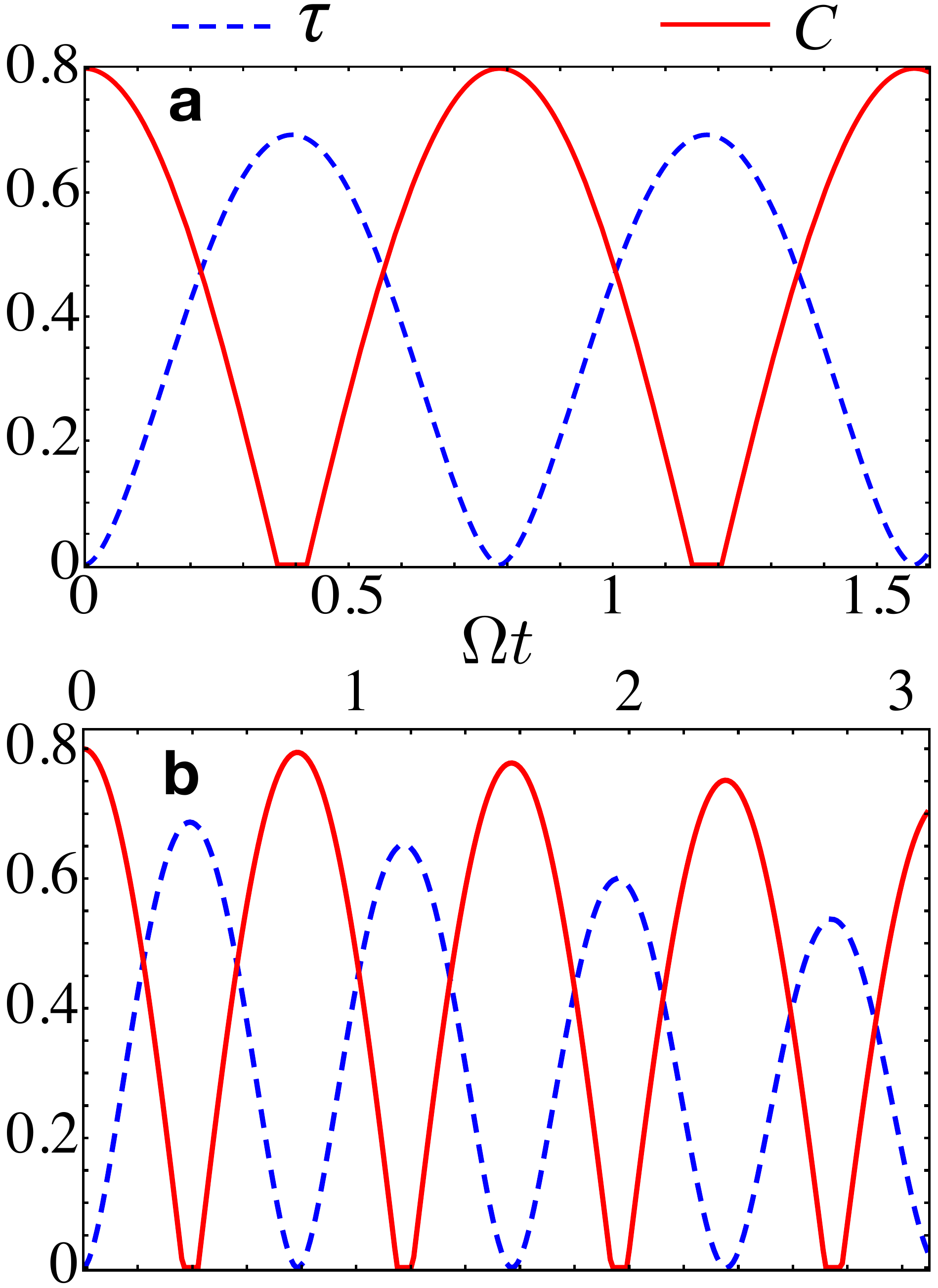}\hspace{1.8 cm}
\includegraphics[width=0.337\textwidth]{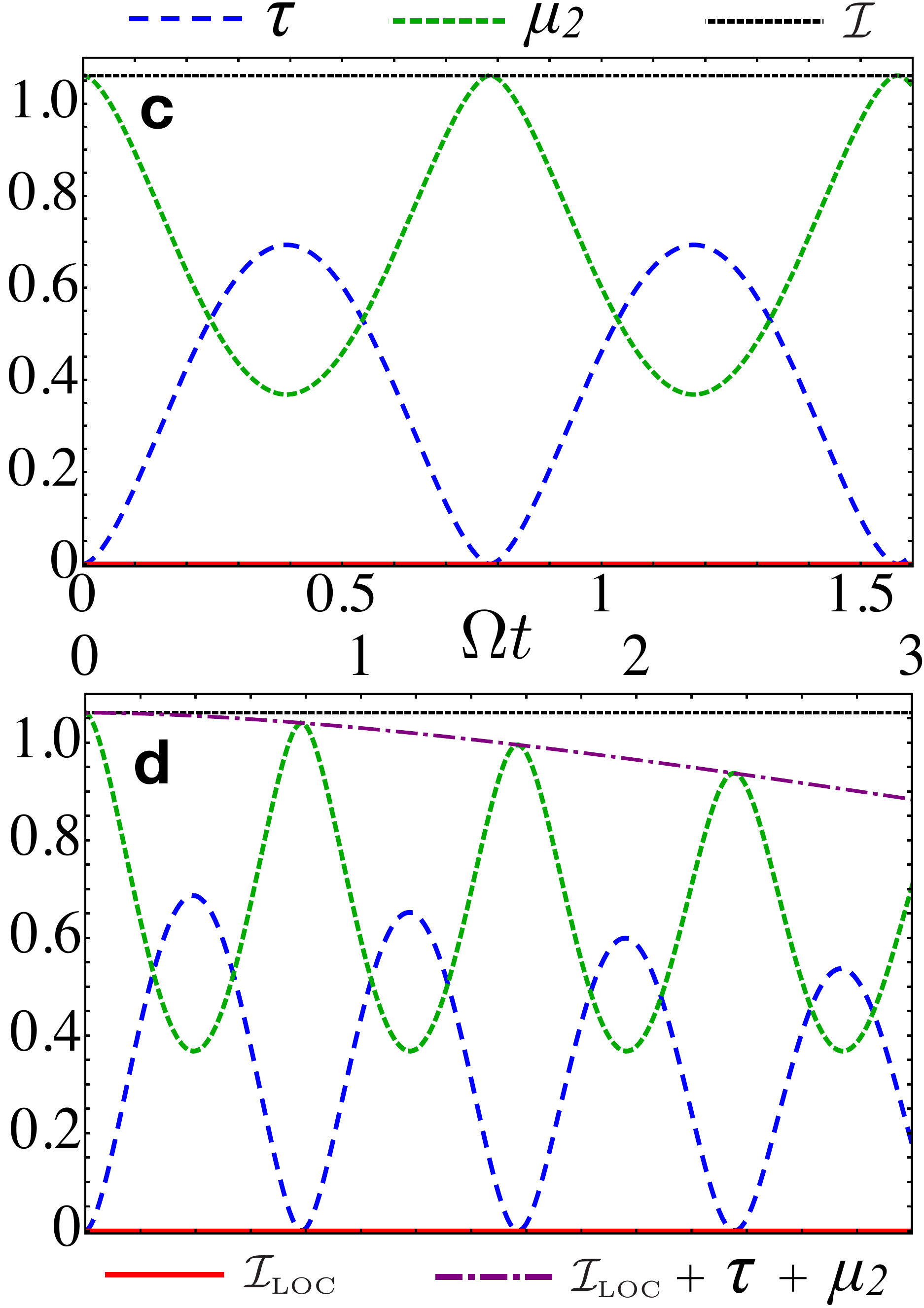}}
\end{center}
\caption{\textbf{Correlation dynamics and information flows.} \textbf{a.} Total tripartite correlations $\tau(\rho_{ABE}(t))$ (blue dashed line) and concurrence $C(\rho_{AB}(t))$ (red solid line) versus $\Omega t$ for the initial Bell-diagonal state $\rho_{AB}^0(1,0.9,1)$ in the case of periodic dynamics ($\sigma\rightarrow 0$, no Gaussian distribution of the Rabi frequency). \textbf{b.} $\tau(\rho_{ABE}(t))$ (blue dashed line) and $C(\rho_{AB}(t))$ (red solid line) for the same initial conditions under decoherent dynamics (Gaussian distribution of the Rabi frequency with $\sigma=0.1\,\Omega$). \textbf{c.} Genuine tripartite correlations $\tau$ (dashed blue line), total state information $\mathcal{I}$ (dotted black line), maximal bipartite correlations $\mu_2$ (green dashed line) and local state information $\mathcal{I}_{\mathrm{LOC}}$ (red solid line) versus $\Omega t$ for the initial Bell-diagonal state $\rho_{AB}^0(1,0.9,1)$ in the case of periodic dynamics ($\sigma\rightarrow 0$). \textbf{d.} The same quantities of panel (c) plotted in the case of decoherent evolution ($\sigma=0.1\,\Omega$). Figures from Ref.~\cite{LeggioPRA}.}
\label{fig:InfoFlux}
\end{figure}

The evolutions of $C(t)=C(\rho_{AB}(t))$ and $\tau(t)=\tau(\rho_{ABE}(t))$, starting from the Bell-diagonal state $\rho_{AB}^0(1,0.9,1)$ considered both in the theoretical model of Sec.~\ref{sec:REF} and in the experiment of Sec.~\ref{sec:Exprevival}, are plotted in Fig. \ref{fig:InfoFlux}(a)-(b) for a direct comparison. The qubit-field interaction reduces the entanglement while correlating the environment with the two-qubit system. Since $B$ and $E$ always remain uncorrelated during the dynamics (as discussed in the previous subsection), correlations in the overall system can only turn into genuine tripartite correlations, as seen from Fig.~\ref{fig:InfoFlux}(a)-(b). When entanglement decreases, genuine tripartite correlations increase, $C$ and $\tau$ showing a time behavior in phase opposition such that the maxima of $\tau$ coincide with the minima of $C$ and viceversa 
\footnote{It is worth to mention here that, if a coherence between Bell states is introduced in the initial two-qubit state, for instance for a state as $\rho_2=\rho_{AB}^0(0.6,0.8,0.3)$ of Eq.~(\ref{initstate}), freezing of genuine tripartite correlations occurs for finite time periods, showing a plateau in correspondence of the plateau of zero entanglement \cite{LeggioPRA}. The discussion of this behavior is out of the scopes of this chapter.}. 

The study of the information fluxes within the system $\{A,B,E\}$ can be conducted by exploiting a relation involving mutual informations $I$, genuine tripartite correlations $\tau$ and von Neumann entropies, which is given by \cite{costaPRA}
\begin{equation}\label{monogamy}
\mathcal{I}=\mathcal{I}_{\mathrm{LOC}}+\tau+\mu_2,
\end{equation}
where $\mathcal{I}=\mathcal{I}(\rho_{ABE})=\ln d-S(\rho_{ABE})$ is the state information of the total tripartite state $\rho_{ABE}$ in the Hilbert space of dimension $d = 2^3 = 8$,
$\mathcal{I}_{\mathrm{LOC}}=\mathcal{I}(\rho_{A})+\mathcal{I}(\rho_{B})+\mathcal{I}(\rho_{E})$ is the total state information locally stored in each part, with $\mathcal{I}(\rho_{i})=\ln d_i-S(\rho_{i})$ ($i=A,B,E$; $d_i=2$), and $\mu_2=\max\{I(\rho_{ij})\}$ is the maximal mutual information over any possible bipartite reduced state $\rho_{ij}$. According to the expression of Eq.~\eqref{monogamy}, local information, tripartite and bipartite correlations thus constitute three containers where the system can store its total information. 
In Fig.~\ref{fig:InfoFlux}(c)-(d) the dynamics of all the quantities involved in Eq.~\eqref{monogamy} is plotted, when the two qubits are initially in the Bell-diagonal state $\rho_{AB}^0(1,0.9,1)$. While in the case of periodic dynamics (closed system) $\mathcal{I}\equiv \mathcal{I}(0)$ is constant, in the decoherent (Gaussian-induced) dynamics the total state information $\mathcal{I}(t)=\tau+\mathcal{I}_{\mathrm{LOC}}+\mu_2$ decays, as shown in panel Fig.~\ref{fig:InfoFlux}(d). A particular feature of the dynamics is that the local state information $\mathcal{I}_{\mathrm{LOC}}$ is constantly zero. The information regarding the total state is always stored in bipartite and (or) tripartite correlations. Precisely, the information is periodically transferred back and forth between bipartite and tripartite correlations.

This behavior is physically understandable by looking at the meaning of the involved quantities within the quantum-classical system under consideration. Since the reduced state of the classical environment is a time invariant maximally mixed state $\rho_E=\frac{1}{2}\sum_{\varphi=\varphi_\pm}|\varphi\rangle\langle \varphi| $, it gives $\mathcal{I}(\rho_{E})=0$. The red line in Fig.~\ref{fig:InfoFlux}(c)-(d) therefore describes the local information $\mathcal{I}_{\mathrm{LOC}}$ due to the two qubits. In this particular case, $\mathcal{I}_{\mathrm{LOC}}=\mathcal{I}(\rho_{A})+\mathcal{I}(\rho_{B})$ is also zero because $\rho_{AB}(t)$ remains a Bell-diagonal state, having maximally mixed marginals for definition and thus $\mathcal{I}(\rho_{A})=\mathcal{I}(\rho_{B})=0$ at any time. 
Genuine tripartite correlations $\tau$, involving all the three parties of the system, represent the information shared among qubit $A$, qubit $B$ and environment $E$. As a consequence, $\mathcal{I}_{\mathrm{LOC}}$ and $\tau$ are two different and non-mixable forms of information stored in the system. A convenient qualitative behavior of the possible fluxes can be provided in a nutshell as \cite{LeggioPRA}
\begin{equation}\label{flows}
\mathcal{I}_{\mathrm{LOC}} \nrightarrow \tau,\quad
\mathcal{I}_{\mathrm{LOC}}\rightleftarrows \mu_2\rightleftarrows \tau,
\end{equation}
which also indicates that local information and bipartite correlations can transform into each other as well as bipartite correlations and genuine tripartite correlations can. Under this scheme, it is clear how in Fig.~\ref{fig:InfoFlux}(c)-(d), where $\mathcal{I}_{\mathrm{LOC}}$ is zero, all the information is stored in correlations which, periodically, change from the bipartite to the tripartite kind, enabling entanglement revivals.

\subsection{Hidden entanglement and lack of classical information}

Another viewpoint about the occurrence of entanglement revivals under local classical noise is based on the local operation and classical communication (LOCC) principle that, if quantum entanglement is restored by means of a local operation after its disappearance, there must be entanglement hidden in the system which does not emerge in the density matrix description of the quantum system \cite{plenioreview}. It is thus useful to introduce the concept of ``hidden entanglement'' \cite{darrigo2012AOP}.

Let us take a bipartite system defined by an ensemble of states ${\cal A}=\{(p_i,|\psi_i\rangle)\}$, for which the statistical distribution of the bipartite pure states $\{|\psi_i\rangle\}$ occurring with probabilities $\{p_i\}$ is known, giving the density matrix $\rho=\sum_i p_i |\psi_i\rangle\langle \psi_i|$. 
The hidden entanglement of the ensemble is defined as \cite{darrigo2012AOP}
\begin{equation}
 E_h({\cal A}) \equiv E_\mathrm{av}({\cal A}) - E(\rho)
 = \sum_i p_i E(|\psi_i\rangle\langle\psi_i|)-E\big(\sum_i p_i |\psi_i\rangle\langle\psi_i|\big),
    \label{eq:MeasHiddenEntang}
\end{equation} 
where ${E}_\mathrm{av}({\cal A})=\sum_i p_i E(|\psi_i\rangle\langle \psi_i|)$ is the average entanglement of the ensemble \cite{Bennett1996,Cohen1998,carvalho2007} and $E(\rho)$ is a convex quantifier of the entanglement of the state (e.g., entanglement of formation) \cite{horodecki2009RMP}. The convexity of $E(\rho)$ ensures that $E_h \geq 0$.
Notice that in the case of a continuous-variable ensemble of the system (see, for instance, the model of low-frequency noise of Sec.~\ref{sec:lowfreqnoise}), the sums become integrals. The hidden entanglement $E_h$ represents the amount of entanglement being unexploitable due to the lack of knowledge of which state of the mixture one is handling. When this classical information is
supplied, such amount of entanglement can be recovered with the only aid of local operations.
Notice that $E_h$ strictly depends on the particular quantum ensemble description of the state of the system. Its role is thus principally relevant in those dynamical situations which, starting from a pure state of the system, univocally determine a physical decomposition in terms of an ensemble of evolved pure states. Interestingly, this situation is always verified when the system is subject to classical noise, which can be treated as an ensemble of local unitaries (random unitaries) applied to the quantum system \cite{darrigo2012AOP}. 

For the case of random external field of Sec.~\ref{sec:REF} with fixed Rabi frequency it is then easy to see that, once fixed the initial two-qubit state $\rho_{AB}(0)$, the physical ensemble is univocally given by
\begin{equation}
{\cal A}(t)=\left\{\left(\frac{1}{2}, (\openone_A\otimes U_{\varphi_+,\Omega} (t)) \rho_{AB}(0)(\openone_A\otimes U_{\varphi_+,\Omega} ^{\dag}(t))\right), \left(\frac{1}{2}, (\openone_A\otimes U_{\varphi_-,\Omega} (t)) \rho_{AB}(0)(\openone_A\otimes U_{\varphi_-,\Omega} ^{\dag}(t))\right)\right\},
\end{equation}
which implies $E_\mathrm{av}({\cal A}(t))=E(\rho_{AB}(0))$ at any times, since the amount of entanglement is invariant under local unitary operations (note that for an initial Bell state, one would have $E_\mathrm{av}({\cal A}(t))=E(\rho_{AB}(0))=1$). Therefore, at times $\bar{t}$ when the entanglement of $\rho_{AB}(\bar{t})$ is zero ($E_f(\rho(\overline{t}))=C(\rho(\overline{t}))=0$), one has a nonzero hidden entanglement $E_h=E_\mathrm{av}({\cal A}(\bar{t}))=E(\rho_{AB}(0))$. The ensemble description points out that this hidden entanglement is due to the lack of knowledge about which local operation is acting on the system. At times $t^\ast$ when this lack of knowledge has no effect, as happens when the two unitaries act as the same operation, entanglement revives reaching its initial value, with $E_f(\rho(t^\ast))=E(\rho_{AB}(0))$ and ${E_h}({\cal A}(t^\ast))=0$ (see the argumentations at the end of above Sec.~\ref{sec:classcontroller}).

For the model with local pulse under low-frequency noise of Sec.~\ref{sec:lowfreqnoise}, where the two-qubit system starts from a Bell state, each realization of $\varepsilon$ gives a pure maximally entangled state forming the ensemble ${\cal A}=\{p(\varepsilon)d\varepsilon,\,\ket{\Psi_\varepsilon(t)}\}$. The average entanglement is $E_\mathrm{av}({\cal A}(t))=1$ at any time. Entanglement decay is due to the lack of classical knowledge on the system $A$-$B$, namely on the random frequency $\varepsilon$. When the pulse is applied (at $t=\overline{t}$), $E_h\approx 1$ and $E_f\approx0$. Entanglement is not destroyed during the evolution but hidden.
After the pulse, this lack of classical knowledge is gradually reduced until $E_h = 0$ and the entanglement reaches its initial value$E_f = 1$ (at $t=2\overline{t}$). The classical information needed to recover entanglement is therefore acquired by means of the local echo pulse.

\subsection{Unifying aspect of the interpretations}

The three mechanisms discussed above which explain the phenomenon of entanglement revivals in classical environment have all a necessary common root: the system dynamics is non-Markovian as defined by the presence of backflows of (classical) information from the environment to the system \cite{breuer2009PRL,breuerRMP}. 

By collecting the main aspects of the interpretations provided so far, the following qualitative considerations can be done: 

\begin{itemize}
\item the classical environment keeps memory of which unitary is acting on the qubit thanks to the occurrence of backflows of classical information; 
\item the periodic transformation of genuine tripartite correlations into two-qubit entanglement is activated by system-environment information fluxes; 
\item local control leads to a partial coherent exchange of information between system and the environment, as also highlighted in the context of discrete qubit dynamics \cite{guo-piilo}, thus allowing the recovery of the hidden entanglement.
\end{itemize}

These considerations can be cast under a general unified view by showing that non-Markovianity of the system dynamics defined by information backflows is the required condition for entanglement revivals to occur in the presence of classical environments.

As previously said, local classical noise can be suitably described as an ensemble of local unitaries \cite{darrigo2012AOP} which make the corresponding dynamical map of the system unital \cite{chruschinskirandom,alickibook}. Under the spectator configuration adopted here typical of the decoherence paradigm (an isolated qubit plus an open qubit interacting with its local environment), it is straightforward to prove that two-qubit entanglement revivals necessarily enable information backflows from the environment to the system and viceversa. In fact, the occurrence of an entanglement non-monotonic evolution within this configuration is just the ground aspect for the non-Markovianity quantifier based on the indivisibility of the dynamical map \cite{rivas2010PRL} which coincides, for unital maps \cite{mannone2012,bylicka2014,haseli2015}, with the quantifier based on distinguishability of quantum states as measured by trace distance \cite{breuer2009PRL}. The latter is then interpreted in terms of information backflows from the environment to the system, where this information can be either quantum (for the case of dissipative quantum environments) or classical (for the case of nondissipative and classical environments) \cite{bylicka2014,breuer2009PRL,walbornPRL,walbornPRA}. Being the dynamical map associated to a classical environment without back-action a unital channel, one finally has the equivalence 
\begin{equation}
\textit{Backflows of Classical Information} \Leftrightarrow \textit{Entanglement Revivals in Classical Environments}.
\end{equation}
Hence, if there is classical information flowing back from the classical environment to the bipartite quantum system in absence of backaction, then entanglement revivals occur; viceversa, if bipartite quantum entanglement revives during the system evolution under a local interaction with a classical environment which does not back react, then system-environment backflows of classical information occur.

\section{Conclusion}\label{sec:conclusion}

In this chapter we have presented an overview about some of the main theoretical and experimental results presenting the phenomenon of revivals of quantum entanglement between two qubits where one qubit only is locally interacting with a classical environment, the other qubit being isolated. This configuration is the simplest one to study the effects of the classical environment on system dynamics and its role in restoring entanglement initially present in the two-qubit system. This has been employed by many theoretical studies considering classical noise made, for instance, of a random external field, pure-dephasing low-frequency noise and random telegraph noise, which are the ones we have explicitly presented here (see Sec.~\ref{sec:TheorPredictions}). Major emphasis has been given to the case of random external field characterized by two random phases, since it constitutes the first instance where a tentative interpretation of the phenomenon of entanglement revivals in classical environments without back-action has been provided \cite{lofranco2012PRA,LeggioPRA}. 

We have then discussed two all-optical experiments reproducing, respectively, the model with a two-phase random external field \cite{LoFrancoNatCom} and the model with dephasing low-frequency noise where a local pulse is applied to make entanglement revive \cite{adeline2014}. Both the experiments confirm the theoretical predictions, presenting direct observations of entanglement revivals (spontaneous or inducted by a local operation) in a classical environment.    

We have also reviewed the three interpretations provided so far for the phenomenon treated in the chapter, all of them supplying responses to the question: where does quantum entanglement go before reappearing during the system dynamics in absence of back-action? This question stands at the basis of the comprehension of the physical mechanisms allowing entanglement revivals under this condition. We notice that for any nondissipative environment, either quantum or classical, back-action is absent and the entanglement revivals should be thus interpreted by the same mechanisms: for instance, this is the case of unital quantum channels such as bit flip, bit-phase flip and phase flip \cite{aaronson2013PRA,aaronson2013NJP,universalfreezing,aliPLA}. The three interpretations respectively rely on three different concepts, which can be summed up as follows: (i) classical environment as a controller keeping a record for what unitary operation acts on the qubit \cite{lofranco2012PRA,LoFrancoNatCom}; (ii) interchange between threepartite correlations and two-qubit entanglement \cite{LeggioPRA}; (iii) hidden entanglement existing in the system which is recoverable by a local operation \cite{darrigo2012AOP}. We have finally shown that these explanations of the phenomenon can be collected under a unified physical aspect, namely the presence of non-Markovianity as defined by the occurrence of backflows of classical information from the classical environment without backaction to the quantum system. In general, all the studies developed so far suggest that information backflows (quantum or classical) are the essential requisite to obtain revivals of quantum features, as also pointed out in Ref.~\cite{PhysRevA.93.042119}, independently of the quantum or classical nature of the environment.

The reviewed results and the argumentations here reported supply a wide insight on the mechanisms underlying the recovery of entanglement in hybrid quantum-classical systems. Such a knowledge can be useful to motivate and boost further studies on the manipulation of hybrid systems for quantum technology \cite{hybridPNAS2015}.

\begin{acknowledgments}
R.L.F. and G.C. acknowledge Diogo Soares Pinto, Felipe Fanchini and Gerardo Adesso for giving them the possibility to contribute to the present book: \textit{Lectures on general quantum correlations and their applications}.  
\end{acknowledgments}

\end{document}